\documentclass[lettersize,journal,review]{IEEEtran}

\usepackage{graphicx}
\usepackage{amssymb,amsmath,epsfig}
\usepackage{algorithm}
\usepackage{algorithmic}
\usepackage{booktabs}
\usepackage{mathrsfs}
\usepackage{color}
\usepackage{textcomp}
\usepackage{bbding}
\usepackage{pifont}
\usepackage{wasysym}
\usepackage{amssymb}
\usepackage{subcaption}
\usepackage{cite}
\usepackage{amsmath,amssymb,amsfonts}
\usepackage{xcolor,colortbl}
\usepackage{overpic}
\usepackage{arydshln}
\usepackage{microtype}
\usepackage{multirow}
\usepackage{makecell}
\usepackage{hhline}
\usepackage[american]{babel}
\usepackage{cuted}
\usepackage{ragged2e}

\definecolor{lightgray}{gray}{.92}
\definecolor{tinygray}{gray}{.96}

\newcommand{\etc}{\textit{etc}}

\usepackage{multirow}
\usepackage{epsfig}
\usepackage{adjustbox}
\usepackage[pagebackref=true,
            breaklinks=true,colorlinks,
            citecolor=cyan,linkcolor=red,
            bookmarks=false]{hyperref}

\ifCLASSINFOpdf
\else
\fi
\usepackage[switch]{lineno}

\begin{document}

\title{Dual-domain Modulation Network for Lightweight Image Super-Resolution}


\author{Wenjie Li,
        Heng Guo\textsuperscript{$\ast$},
        Yuefeng Hou,
        Guangwei Gao,
        and Zhanyu Ma
 \IEEEcompsocitemizethanks{\IEEEcompsocthanksitem This work was supported by Beijing-Tianjin-Hebei Basic Research Funding Program No. F2024502017, National Natural Science Foundation of China (Grant No. 62472044, U24B20155, 62225601, U23B2052), Hebei Natural Science Foundation Project No. 242Q0101Z, Beijing Natural Science Foundation Project No. L242025. ($\ast$: Corresponding author) 
}
\IEEEcompsocitemizethanks{\IEEEcompsocthanksitem Wenjie Li, Heng Guo, and Zhanyu Ma are with the Pattern Recognition and Intelligent System Laboratory, School of Artificial Intelligence, Beijing University of Posts and Telecommunications (BUPT), Beijing 100080, China (e-mail: \{cswjli, guoheng, mazhanyu\}@bupt.edu.cn). 
}
\IEEEcompsocitemizethanks{\IEEEcompsocthanksitem Yuefeng Hou is with the School of Microelectronics, Tianjin University (TJU), Tianjin 300072, China (e-mail: houyuefeng@tju.edu.cn).
}
\IEEEcompsocitemizethanks{\IEEEcompsocthanksitem Guangwei Gao is with the Institute of Advanced Technology, Nanjing University of Posts and Telecommunications (NJUPT), Nanjing 210046, China (e-mail: csggao@gmail.com).}
}

\markboth{IEEE TRANSACTIONS ON MULTIMEDIA}%
{Shell \MakeLowercase{\textit{et al.}}: A Sample Article Using IEEEtran.cls for IEEE Journals}

\maketitle

\begin{abstract}
Lightweight image super-resolution (SR) aims to reconstruct high-resolution images from low-resolution images under limited computational costs. We find that existing frequency-based SR methods cannot balance the reconstruction of overall structures and high-frequency parts. Meanwhile, these methods are inefficient for handling frequency features and unsuitable for lightweight SR. In this paper, we show that introducing both wavelet and Fourier information allows our model to consider both high-frequency features and overall SR structure reconstruction while reducing costs. Specifically, we propose a Dual-domain Modulation Network that integrates both wavelet and Fourier information for enhanced frequency modeling. Unlike existing methods that rely on a single frequency representation, our design combines wavelet-domain modulation via a Wavelet-domain Modulation Transformer (WMT) with global Fourier supervision, enabling complementary spectral learning well-suited for lightweight SR. Experimental results show that our method achieves a comparable PSNR to SRFormer~\cite{zhou2023srformer} and MambaIR~\cite{guo2025mambair} while with less than 50\% and 60\% of their FLOPs and achieving inference speeds 15.4× and 5.4× faster, respectively, demonstrating the effectiveness of our method on SR quality and lightweight. Code link: \url{https://github.com/24wenjie-li/DMNet}
\end{abstract}




\IEEEpeerreviewmaketitle

\section{Introduction}\label{introduction}
\begin{figure}[t]
\begin{overpic}[width=0.99\linewidth]{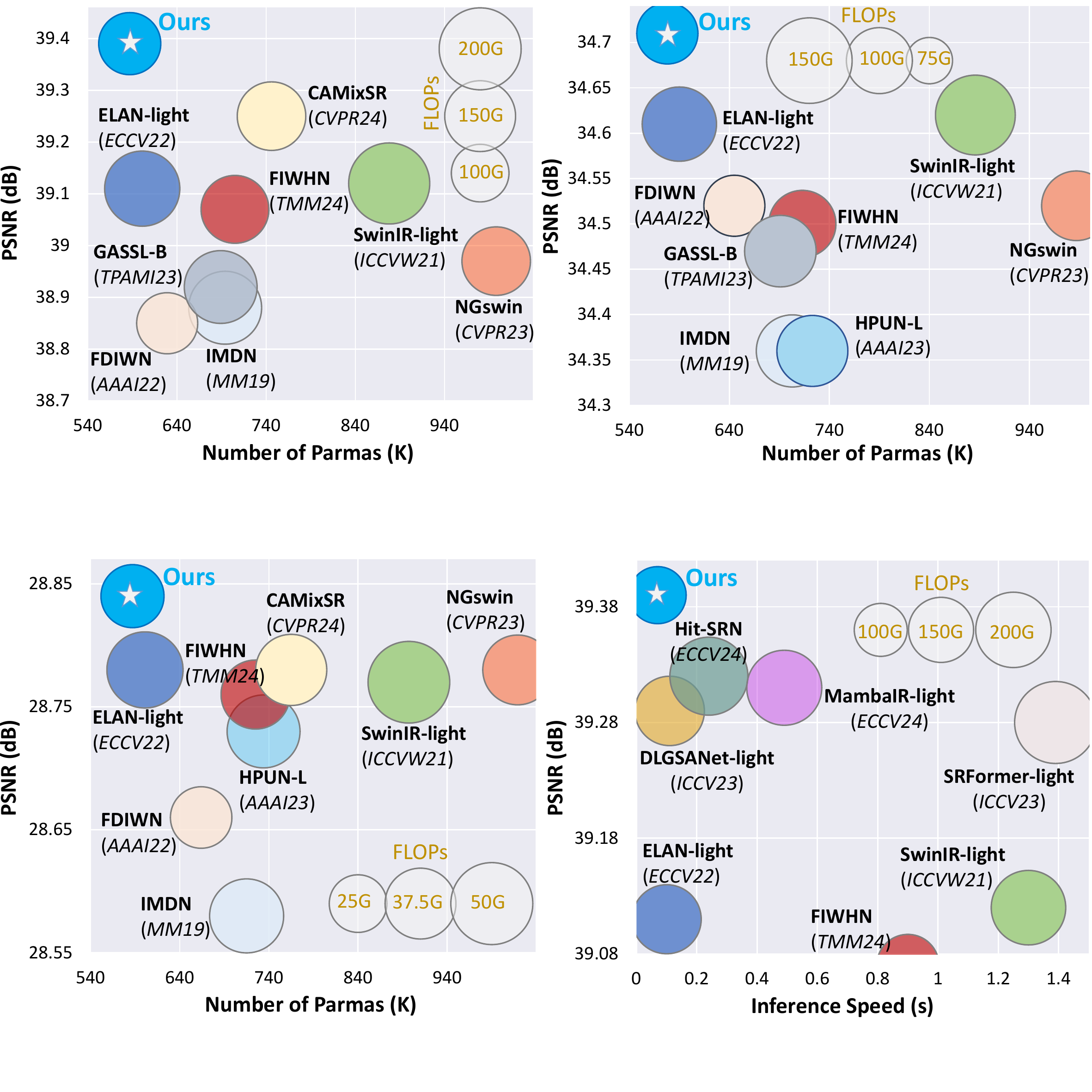}
\put(0.8,53){\color{black}{\fontsize{7pt}{1pt}\selectfont (a) Tradd-off on $\times 2$ Manga109~\cite{matsui2017sketch} test set.}}
\put(54.8,53){\color{black}{\fontsize{7pt}{1pt}\selectfont (b) Tradd-off on $\times 3$ Set5~\cite{bevilacqua2012low} test set.}}

\put(2.8,2){\color{black}{\fontsize{7pt}{1pt}\selectfont (c) Tradd-off on $\times 4$ Set14~\cite{zeyde2012single} test set.}}
\put(50.8,2){\color{black}{\fontsize{7pt}{1pt}\selectfont (d) Tradd-off on $\times 2$ Manga109~\cite{matsui2017sketch} test set.}}

\put(11.1,20){\color{black}{\fontsize{4.5pt}{1pt}\selectfont \cite{gao2022feature}}}
\put(17.4,16.7){\color{black}{\fontsize{4.5pt}{1pt}\selectfont \cite{hui2019lightweight}}}
\put(27.8,28.5){\color{black}{\fontsize{4.5pt}{1pt}\selectfont \cite{sun2023hybrid}}}
\put(10.1,29.8){\color{black}{\fontsize{4.5pt}{1pt}\selectfont \cite{zhang2022efficient}}}
\put(35.8,28.4){\color{black}{\fontsize{4.5pt}{1pt}\selectfont \cite{liang2021swinir}}}
\put(40.8,39.99){\color{black}{\fontsize{4.5pt}{1pt}\selectfont \cite{choi2023n}}}
\put(31.6,44.4){\color{black}{\fontsize{4.5pt}{1pt}\selectfont \cite{wang2024camixersr}}}
\put(19.1,35.8){\color{black}{\fontsize{4.5pt}{1pt}\selectfont \cite{li2022efficient}}}

\put(66.9,21.4){\color{black}{\fontsize{4.5pt}{1pt}\selectfont \cite{zhang2022efficient}}}
\put(85.2,16.6){\color{black}{\fontsize{4.5pt}{1pt}\selectfont \cite{liang2021swinir}}}
\put(90.7,23.9){\color{black}{\fontsize{4.5pt}{1pt}\selectfont \cite{zhou2023srformer}}}
\put(63.7,26.15){\color{black}{\fontsize{4.5pt}{1pt}\selectfont \cite{li2023dlgsanet}}}
\put(87.1,36.0){\color{black}{\fontsize{4.5pt}{1pt}\selectfont \cite{guo2025mambair}}}
\put(67.6,41.8){\color{black}{\fontsize{4.5pt}{1pt}\selectfont \cite{zhang2025hit}}}
\put(79.4,15.2){\color{black}{\fontsize{4.5pt}{1pt}\selectfont \cite{li2022efficient}}}

\put(66.8,68.4){\color{black}{\fontsize{4.5pt}{1pt}\selectfont \cite{hui2019lightweight}}}
\put(84.1,69.7){\color{black}{\fontsize{4.5pt}{1pt}\selectfont \cite{sun2023hybrid}}}
\put(94.9,72.6){\color{black}{\fontsize{4.5pt}{1pt}\selectfont \cite{choi2023n}}}
\put(66.99,76.2){\color{black}{\fontsize{4.5pt}{1pt}\selectfont \cite{wang2023global}}}
\put(63.2,82.1){\color{black}{\fontsize{4.5pt}{1pt}\selectfont \cite{gao2022feature}}}
\put(93.0,84.0){\color{black}{\fontsize{4.5pt}{1pt}\selectfont \cite{liang2021swinir}}}
\put(74.17,88.5){\color{black}{\fontsize{4.5pt}{1pt}\selectfont \cite{zhang2022efficient}}}
\put(82.25,79.5){\color{black}{\fontsize{4.5pt}{1pt}\selectfont \cite{li2022efficient}}}

\put(14.1,66.7){\color{black}{\fontsize{4.5pt}{1pt}\selectfont \cite{gao2022feature}}}
\put(24.1,67.6){\color{black}{\fontsize{4.5pt}{1pt}\selectfont \cite{hui2019lightweight}}}
\put(43.5,67.4){\color{black}{\fontsize{4.5pt}{1pt}\selectfont \cite{choi2023n}}}
\put(34.5,73.6){\color{black}{\fontsize{4.5pt}{1pt}\selectfont \cite{liang2021swinir}}}
\put(15.2,76.5){\color{black}{\fontsize{4.5pt}{1pt}\selectfont \cite{wang2023global}}}
\put(25.5,78.9){\color{black}{\fontsize{4.5pt}{1pt}\selectfont \cite{li2022efficient}}}
\put(29.5,86.9){\color{black}{\fontsize{4.5pt}{1pt}\selectfont \cite{wang2024camixersr}}}
\put(17.3,88.9){\color{black}{\fontsize{4.5pt}{1pt}\selectfont \cite{zhang2022efficient}}}

\end{overpic}
   \caption{Efficiency trade-offs between our method and existing methods. Our method achieves a better balance in terms of PSNR, model size, and inference speed than existing methods.}
\label{fig:teaser}
\end{figure}

\IEEEPARstart{I}{MAGE} super-resolution (SR) aims at reconstructing high-resolution (HR) images from low-resolution (LR) images, which has achieved impressive performance in medical imaging~\cite{georgescu2023multimodal}, remote sensing~\cite{xiao2023ediffsr}, security surveillance~\cite{jiang2020dual}, and text enhancement~\cite{noguchi2024scene}.
Most SR methods~\cite{lim2017enhanced,dai2019second} require high computational costs, limiting their application in devices with limited resources. As shown in Fig.~\ref{fig:teaser}, we focus on the lightweight SR task, aiming to improve existing SR performance while reducing model size and accelerating inference.

\begin{figure*}[ht]
\begin{overpic}[width=0.99\linewidth]{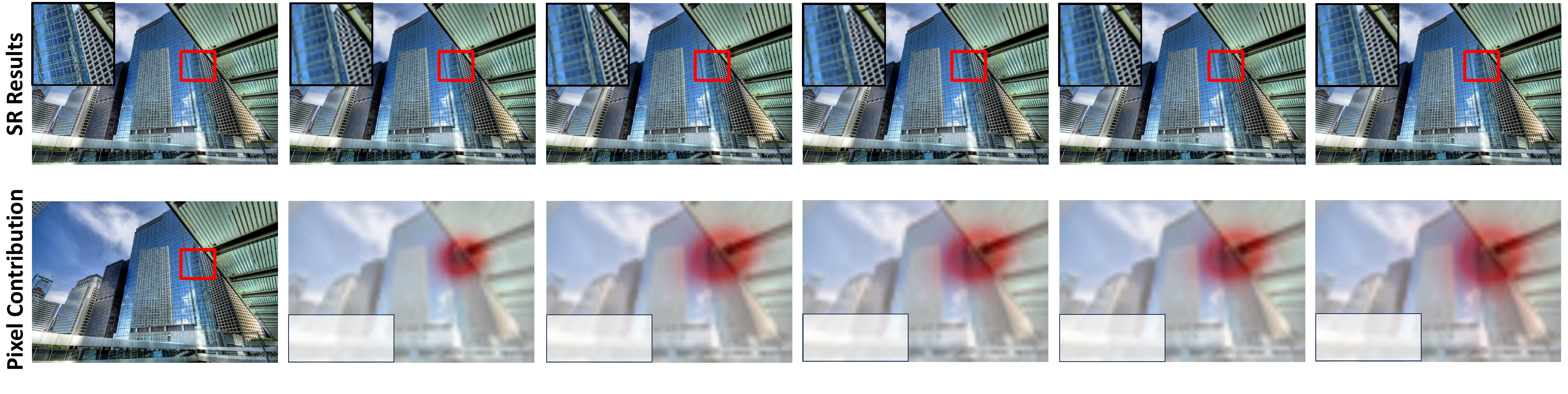}
\put(5.3,0.3){\color{black}{\fontsize{7.2pt}{1pt}\selectfont Ground Truth}}
\put(7.8,-1.6){\color{black}{\fontsize{7.2pt}{1pt}\selectfont (HR)}}

\put(19.8,0.3){\color{black}{\fontsize{7.2pt}{1pt}\selectfont Spatial Reconstruction}}
\put(20.9,-1.6){\color{black}{\fontsize{7.2pt}{1pt}\selectfont Spatial Supervision}}

\put(36.0,0.3){\color{black}{\fontsize{7.2pt}{1pt}\selectfont Spatial Reconstruction}}
\put(37.0,-1.6){\color{black}{\fontsize{7.2pt}{1pt}\selectfont Fourier Supervision}}

\put(52.6,0.3){\color{black}{\fontsize{7.2pt}{1pt}\selectfont Fourier Reconstruction}}
\put(53.6,-1.6){\color{black}{\fontsize{7.2pt}{1pt}\selectfont Fourier Supervision}}

\put(68.7,0.3){\color{black}{\fontsize{7.2pt}{1pt}\selectfont Wavelet Reconstruction}}
\put(69.7,-1.6){\color{black}{\fontsize{7.2pt}{1pt}\selectfont Wavelet Supervision}}

\put(85.0,0.3){\color{black}{\fontsize{7.pt}{1pt}\selectfont Wavelet Reconstruction}}
\put(84.1,-1.6){\color{black}{\fontsize{7.2pt}{1pt}\selectfont Fourier Supervision \textbf{(Ours)}}}


\put(19.5,3.0){\color{black}{\fontsize{8pt}{1pt}\selectfont DI: 4.2}}
\put(35.4,3.0){\color{black}{\fontsize{8pt}{1pt}\selectfont DI: 9.7}}
\put(51.8,3.0){\color{black}{\fontsize{8pt}{1pt}\selectfont DI: 11.9}}
\put(68.2,3.0){\color{black}{\fontsize{8pt}{1pt}\selectfont DI: 10.8}}
\put(84.6,3.0){\color{black}{\fontsize{8pt}{1pt}\selectfont \textbf{DI: 16.4}}}


\put(3.4,13.2){\color{black}{\fontsize{8pt}{1pt}\selectfont PSNR/SSIM/Parmas}}
\put(20.0,13.2){\color{black}{\fontsize{8pt}{1pt}\selectfont 24.61/0.7518/594K}}
\put(36.2,13.2){\color{black}{\fontsize{8pt}{1pt}\selectfont 25.00/0.7718/594K}}
\put(52.6,13.2){\color{black}{\fontsize{8pt}{1pt}\selectfont 25.01/0.7717/598K}}
\put(68.9,13.2){\color{black}{\fontsize{8pt}{1pt}\selectfont 24.90/0.7563/588K}}
\put(85.8,13.2){\color{black}{\fontsize{8pt}{1pt}\selectfont \textbf{25.18/0.7738/587K}}}

\end{overpic}
\vspace{4.5mm}
   \caption{By reconstructing features in the wavelet domain and supervising the process in the Fourier domain, our method recovers precise textures and achieves the highest PSNR. Meanwhile, compared to existing methods~\cite{sun2022shufflemixer,kong2023efficient,zou2022joint}—all adapted to use the same spatial module as ours—our method achieves a higher diffusion index (DI)~\cite{gu2021interpreting} score, which quantifies the pixel contribution area, demonstrating its ability to capture a broader and more relevant pixel range for reconstruction.}
\label{fig:movitation}
\vspace{-0mm}
\end{figure*}

Existing lightweight SR methods~\cite{liang2021swinir,zhang2022efficient,wang2024camixersr} primarily operate in the \emph{spatial domain}, relying solely on spatial-domain reconstruction and supervision. This design inherently limits the effective receptive field during inference, as illustrated in the second column of Fig.\ref{fig:movitation}. In contrast, LAM \cite{gu2021interpreting} suggests that leveraging a broader contextual pixel range is beneficial for improving SR, and diffusion index (DI) \cite{gu2021interpreting} serves as a quantitative indicator of this effect. To compensate for the limited pixel utilization, these methods often resort to stacking numerous blocks, which leads to increased computational cost and inefficiency. For example, SwinIR-light~\cite{liang2021swinir} and SRFormer-light~\cite{zhou2023srformer} have nearly 900K parameters and more than 1.3 seconds of inference speed at the scale of $\times 2$, respectively. Therefore, lightweight SR still remains an open problem.

In this paper, we handle the lightweight SR with the spatial and \emph{frequency domain} reconstruction. Our insight comes from recent SR methods~\cite{sun2022shufflemixer,sun2023spatially,zou2022joint,li2023fsr} that expand the range of pixel utilization with additional frequency information, typically in the Fourier or wavelet domains, to improve SR performance. However, as shown in the third to fifth columns of Fig.~\ref{fig:movitation}, compared to spatial-based methods, though these methods expand the range of pixel utilization with fewer params, we also observe these methods face challenges in balancing high-frequency textures and overall structures (PSNR/SSIM). Specifically, Fourier domain-based methods~\cite{sun2022shufflemixer,sun2023spatially,kong2023efficient,jiang2023fabnet} help regulate the global frequency distribution, optimizing the overall structure and leading to high PSNR values. However, it fails to effectively emphasize high-frequency features, which results in blurred high-frequency textures. Wavelet domain-based methods~\cite{zou2022joint,li2023fsr} facilitate emphasizing high-frequency features of images, resulting in clear high-frequency textures. However, the inconsistency of different frequency subbands in the direction of gradient descent is large and unfavorable for controlling losses, thus limiting overall structures.

To leverage frequency analysis for lightweight SR and address the challenge of balancing high-frequency textures and overall structures, we reconstruct features in the wavelet domain and apply supervision in the Fourier domain. Specifically, wavelet transforms effectively decompose high- and low-frequency subbands, facilitating the modeling of their relationships, while Fourier supervision provides an efficient means to regulate the global frequency distribution. As shown in the sixth column of Fig.~\ref{fig:movitation}, compared to existing frequency-based methods, our method expands the range of pixel utilization (see DI score) while avoiding blurred high-frequency textures or inaccurate overall structures with the least Params.

Based on the above observations, we propose a Dual-domain Modulation Network (DMNet) for lightweight SR. Our DMNet contains two modules: Wavelet-domain Modulation Transformer (WMT) and Spatial-domain Modulation Transformer (SMT), which modulate the wavelet and spatial domain information separately, which utilize self-attention to separately capture the correlations among different wavelet frequency and spatial domain features. Meanwhile, we separately utilize Fourier loss and image reconstruction loss to supervise frequency and spatial domain reconstruction. Our design ensures the overall similarity of SR images with ground truth while highlighting high-frequency features absent in LR images. Additionally, WMT combines self-attention and dynamic convolution to extract wavelet-domain features, capturing long-range dependencies and adapting to various frequency components, enabling efficient processing of frequency-domain features. As depicted in Fig.~\ref{fig:teaser} (a), (b), (c), and (d) our method achieves the highest PSNR with fewer Params and FLOPs compared to existing methods, and our method delivers the highest PSNR with minimal FLOPs and the fastest inference speed compared to recent Transformer- and Mamba-based methods. For example, on ×2 Manga109~\cite{matsui2017sketch} test set, our DMNet achieves 0.07 dB higher than Hit-SNG~\cite{zhang2025hit} while using only 56\% of its Parmas and 54\% of its FLOPs—and surpasses MambaIR~\cite{guo2025mambair} by 0.08 dB with 35\% fewer Parmas and 42\% fewer FLOPs.

In summary, the contributions of this paper are as follows:

\begin{itemize}
  \item We find hybrid frequency domain utilization, combining wavelet features with Fourier supervision, effectively balances overall structure and high-frequency reconstruction.
  \item We propose a WMT based on wavelet transform, which can highlight the relationship between high- and low-frequency contents, further improving SR texture accuracy.
  \item Extensive experiments demonstrate that our DMNet can outperform existing lightweight SR methods in synthetic and real datasets with fewer Params and FLOPs.
\end{itemize}


\begin{table}[t]
\tiny
\setlength\tabcolsep{1pt}
\centering
\vspace{0mm}
    \caption{Comparison of our method with existing frequency SR methods, where CNN or Transformer denotes the structure for utilizing frequency-domain information. Our method combines the advantages of hybrid frequency domains and introduces a new lightweight frequency-based Transformer module.}
\label{tab: differnet}
\resizebox{0.48\textwidth}{!}{
\begin{tabular}{l||c|c}
\toprule
\rowcolor{lightgray}
    \rowcolor{lightgray}
    \multicolumn{1}{l||}{\multirow{-1}{*}{Frequency Type}} 
    & \multicolumn{1}{c|}{\multirow{-1}{*}{CNN Structure}}
    & \multicolumn{1}{c}{\multirow{-1}{*}{Transformer Structure}}

    \\ 
    \hline

    \makecell{Fourier domain \\ (Blur textures)}  & \makecell{SFMNet \cite{zheng2025smfanet}, \\ FDSR \cite{xu2024fdsr}, \etc \\ } & \makecell{FFTFormer~\cite{kong2023efficient}, SFANet~\cite{guo2023spatial}, \etc} \\ 
    \hline
    \makecell{Wavelet domain  \\ (Poor PSNR)} & \makecell{JWSGN~\cite{zou2022joint}, \\ WFEN~\cite{li2024efficient}, \etc}  & FSR~\cite{li2023fsr}, SODA-SR~\cite{ai2024uncertainty}, \etc\\ 
    \hline
    \makecell{Wavelet+Fourier} & -  & \makecell{\textbf{Ours} (Clear textures and high PSNR)} \\ 

    \bottomrule
\end{tabular}}
\end{table}
\hspace{-0mm}

\begin{figure*}[ht]
\begin{overpic}[width=0.99\linewidth]{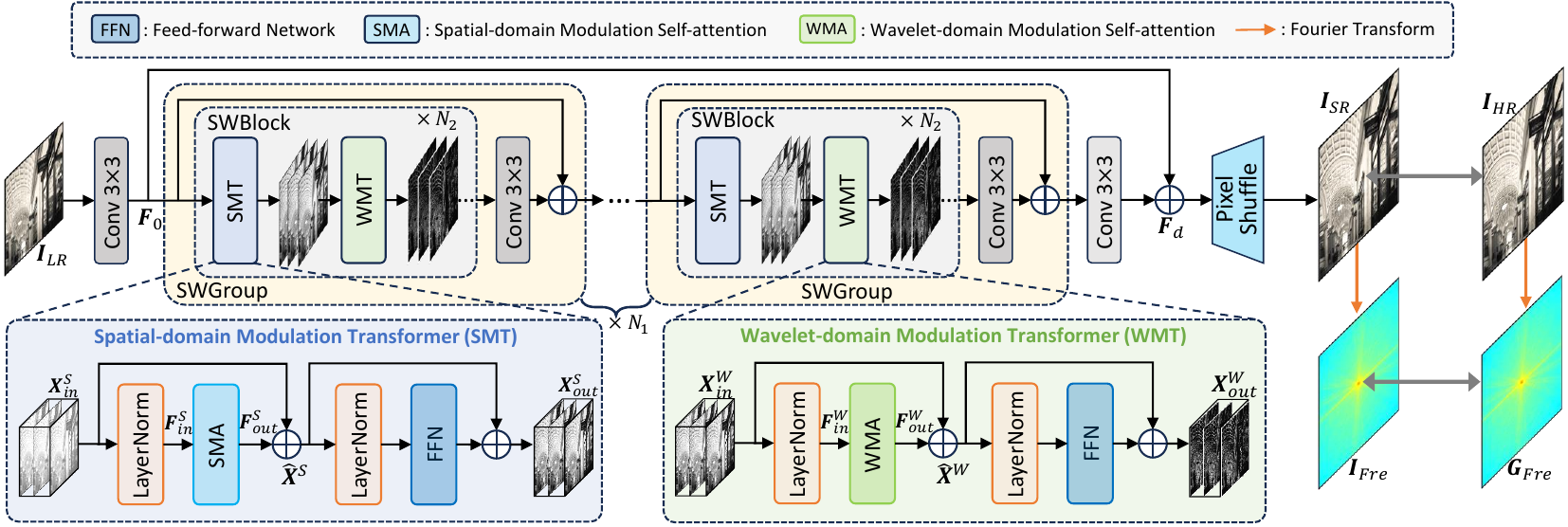}
\put(89.3,24.5){\color{black}{\fontsize{8pt}{1pt}\selectfont \hbox{${\mathcal{L}_{pixel}}$}}}
\put(89.3,11.3){\color{black}{\fontsize{8pt}{1pt}\selectfont \hbox{${\mathcal{L}_{fre}}$}}}
\end{overpic}
\vspace{-0mm}
   \caption{Overview of our method, which utilizes dual modulation in spatial and frequency domains for lightweight SR. Our method comprises \hbox{${N_1}$} Spatial-Wavelet Groups (SWGroup), each SWGroup comprises \hbox{${N_2}$} Spatial-Wavelet Blocks (SWBlock).}
\label{fig:main}
\vspace{-0mm}
\end{figure*}

\vspace{-3mm}
\section{Related Work}~\label{Related}
\vspace{-8mm}
\subsection{Lightweight SR Methods}
To promote realistic adoption~\cite{li2023survey,liu2025causal,li2025saratr}, lightweight SR~\cite{li2024systematic} has grown significantly. IMDN \cite{hui2019lightweight} introduces multilevel feature distillation to enhance models. HDN \cite{jiang2020hierarchical} facilitates feature representation while saving memory through hierarchical matrix structure design. DDistill-SR  \cite{wang2023ddistill} incorporates reparameterization on top of information distillation to further improve efficiency. DRSAN \cite{park2021dynamic} leverages dynamic residual attention to limit complexity by adaptively adjusting its structure based on input statistics. FDIWN \cite{gao2022feature} and FIWHN \cite{li2022efficient} improve efficiency by mitigating the activation function's negative impact. To extract long-range features, SwinIR \cite{liang2021swinir} introduces window-based Transformers to learn globally to address the demand for long-range modeling features. ELAN \cite{zhang2022efficient} accelerates the model by sharing weights across self-attention. LBNet \cite{gao2022feature} and CFIN \cite{li2023cross} design efficient architectures to integrate Transformer with CNN. NGswin \cite{choi2023n} overcome the disadvantage of the limited receptive field of window-based Transformer. CAMixSR \cite{wang2024camixersr} divides and conquers different area features using CNN and Transformer. Mamba-based methods\cite{xiao2024frequency,guo2025mambair} reduce computational complexity via linear scanning attention. However, the above methods do not utilize frequency domain features. The difference between ours and existing SR methods is shown in TABLE~\ref{tab: differnet}.


\subsection{Frequency-based SR Methods}
Due to their ability to effectively capture different details, frequency-based SR methods have been developed. Specifically, WDRN \cite{xin2020wavelet} achieves SR by learning different wavelet coefficients. DFSA \cite{magid2021dynamic} focuses on utilizing high-frequency information to compensate for textures. JWSGN \cite{zou2022joint} enhances performance by separately processing features in different wavelet domains. FSR \cite{li2023fsr} accelerates networks by considering data characteristics in different wavelet domains. DDL \cite{liu2023spectral} improves high-frequency parts by integrating Bayesian methods and frequency inference. FDSR \cite{xu2024fdsr} facilitates various features by constructing different frequency networks. Shufflemixer~\cite{sun2022shufflemixer}, SAFMN~\cite{guo2023spatial}, and EFHSSR~\cite{song2024efficient} additionally use Fourier supervision to improve SR. FSAS~\cite{kong2023efficient} and DF-MSRN~\cite{ju2025towards} integrate Fourier domain information to explore global features for restoration. FMSR \cite{xiao2024frequency} enhances performance by integrating a frequency selection module and spatial-frequency fusion. However, they do not combine the respective strengths of wavelet and Fourier domains to facilitate SR. In contrast, our method combines the advantages of a hybrid frequency domain and presents improved self-attention to make it suitable for exploring various frequency features in the wavelet domain.

%



\section{Proposed Method}~\label{Method}
As shown in Fig.~\ref{fig:main}, an LR image is reconstructed into an accurate SR image in the spatial and frequency domain under the supervision of image reconstruction loss \hbox{${\mathcal{L}_{pixel}}$} and frequency loss \hbox{${\mathcal{L}_{fre}}$}. Our DMNet mainly utilizes our proposed SMT and WMT to modulate the spatial and frequency domain separately.  Specifically, for a LR input \hbox{${{\boldsymbol{I}_{LR}}}$~$\in$~$\mathbb{R}^{H\times W \times 3}$}, our method aims to obtain a spatial domain SR output \hbox{${{\boldsymbol{I}_{SR}}}$~$\in$~$\mathbb{R}^{sH\times sW \times 3}$} and a frequency domain SR output \hbox{${{\boldsymbol{I}_{Fre}}}$~$\in$~$\mathbb{R}^{sH\times sW \times 3}$}, where $s$ represents the SR scale factor. Specifically, we first extracts its shallow feature \hbox{${{\boldsymbol{F}_{0}}}$~$\in$~$\mathbb{R}^{H\times W \times C}$}. Next, \hbox{${{\boldsymbol{F}_{0}}}$} is gradually converted to the depth feature \hbox{${{\boldsymbol{F}_{d}}}$~$\in$~$\mathbb{R}^{H\times W \times C}$} through \hbox{${N_1}$} cascaded SWGroups, where one SWGroup contains \hbox{${N_2}$} SWBlocks. Finally, Pixel-Shuffle performs convolution, channel and spatial dimension rearrangement operations on \hbox{${{\boldsymbol{F}_{d}}}$} to output an SR image \hbox{${{\boldsymbol{I}_{SR}}}$}. Then we acquire the frequency domain SR output \hbox{${{\boldsymbol{I}_{Fre}}}$} through Fourier transform. In the following, we introduce our Wavelet-domain Modulation Transformer and Spatial-domain Modulation Transformer in detail, respectively.

\begin{figure*}[ht]
\begin{overpic}[width=0.99\linewidth]{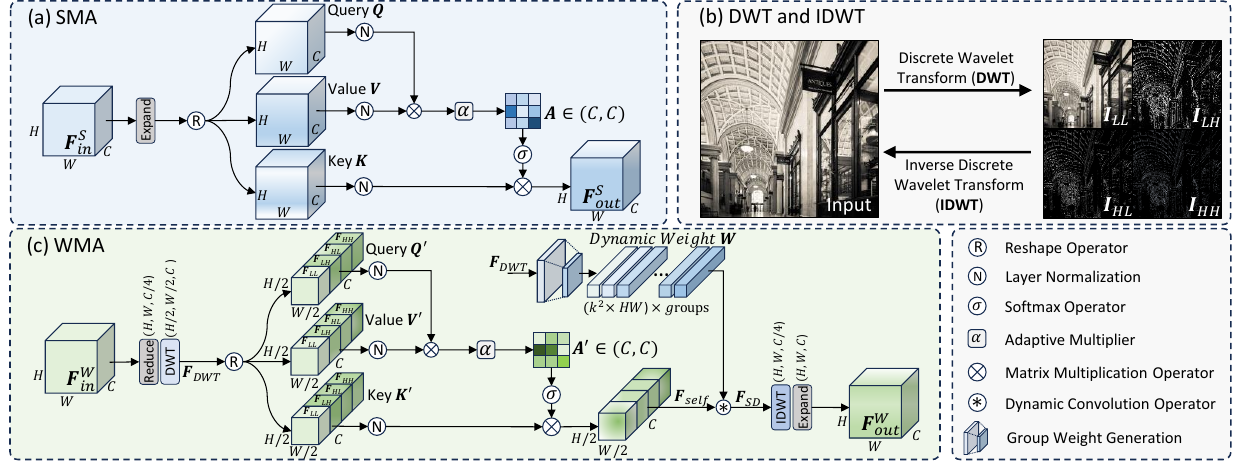}
\end{overpic}
\vspace{-0mm}
   \caption{An illustration of (a) our Spatial-domain Modulation Self-attention (SMA), (b) Discrete Wavelet Transform (DWT) and Inverse Discrete Wavelet Transform  (IDWT) over an image, (c) our Wavelet-domain Modulation Self-attention (WMA).}
\label{fig:Transformer}
\end{figure*}

\subsection{Wavelet-domain Modulation Transformer}
Recent works~\cite{xiao2025multi, jiang2024fmrnet,peng2024lightweight,peng2025boosting,li2025fouriersr} show frequency priors with great potential that integrate wavelet or frequency-aware designs to enhance structural representation. Inspired by this, we adopt DWT to exploit hierarchical frequency cues in a lightweight manner. Specifically, as shown in Fig.~\ref{fig:main}, we propose a WMT to explore the potential relationship between features in different frequency subbands. As shown in Fig.~\ref{fig:Transformer} (c), the core of WMT is to explore the potential relationship between high-frequency and low-frequency features on channels obtained after discrete wavelet transform through our wavelet-domain modulation self-attention. Additionally, a feed-forward network~\cite{zamir2022restormer} ($\mathcal{FFN}$) aggregates the obtained frequency domain features. For an input feature \hbox{$\boldsymbol{X}_{in}^{W}$~$\in$~$\mathbb{R}^{H\times W \times C}$} in Fig.~\ref{fig:main}, the process of obtaining output \hbox{$\boldsymbol{X}_{out}^{W}$~$\in$~$\mathbb{R}^{H\times W \times C}$} can be described as:
\begin{equation}
\hat{\boldsymbol{X}}^{W}   = \mathcal{WMA}\left( {\mathcal{LN}\left( {{\boldsymbol{X}_{in}^{W}}} \right)} \right) + \boldsymbol{X}_{in}^{W} ,
\end{equation}
\begin{equation}
\boldsymbol{X}_{out}^{W}   = \mathcal{FFN}\left( {\mathcal{LN}\left( \hat{\boldsymbol{X}}^{W} \right)} \right) + \hat{\boldsymbol{X}}^{W} ,
\end{equation}
where \hbox{$\mathcal{WMA}$} is our wavelet-domain modulation self-attention (WMA). $\mathcal{LN}$ is layer normalization.

\paragraph{Discrete Wavelet Transform (DWT)} As shown in Fig.~\ref{fig:Transformer} (b), DWT can losslessly convert input \hbox{$\boldsymbol{I}$~$\in$~$\mathbb{R}^{H\times W\times C }$} into one low-frequency subband \hbox{${\boldsymbol{I}_{LL}}$~$\in$~$\mathbb{R}^{\frac{H}{2} \times \frac{W}{2}\times C }$} and three high-frequency subbands \hbox{$\{ {\boldsymbol{I}_{LH}},{\boldsymbol{I}_{HL}},{\boldsymbol{I}_{HH}}\} $~$\in$~$\mathbb{R}^{\frac{H}{2} \times \frac{W}{2}\times C }$}, where \hbox{${\boldsymbol{I}_{LL}}$} reflect coarse-grained basic structure of \hbox{$\boldsymbol{I}$}, and \hbox{$\{ {\boldsymbol{I}_{LH}},{\boldsymbol{I}_{HL}},{\boldsymbol{I}_{HH}}\} $} reflect fine-grained texture details of \hbox{$\boldsymbol{I}$}. Specifically, DWT first encodes \hbox{$\boldsymbol{I}$} into two subbands \hbox{$\{ {\boldsymbol{I}_{L}},{\boldsymbol{I}_{H}}\} $~$\in$~$\mathbb{R}^{\frac{H}{2} \times W\times C }$} using a high-pass filter \hbox{${\mathcal{F}_{H}} = \left( {\frac{1}{{\sqrt 2 }},\frac{{ - 1}}{{\sqrt 2 }}} \right)$} and a low-pass filter \hbox{${\mathcal{F}_{L}} = \left( {\frac{1}{{\sqrt 2 }},\frac{1}{{\sqrt 2 }}} \right)$} along rows. Then, \hbox{${\mathcal{F}_{L}}$} and \hbox{${\mathcal{F}_{H}}$} are encoded along the columns for \hbox{${\boldsymbol{I}_{L}}$} and \hbox{${\boldsymbol{I}_{H}}$} to obtain four wavelet subbands \hbox{$\{ {\boldsymbol{I}_{LL}},{\boldsymbol{I}_{LH}},{\boldsymbol{I}_{HL}},{\boldsymbol{I}_{HH}}\} $~$\in$~$\mathbb{R}^{\frac{H}{2} \times \frac{W}{2}\times C }$}. This process is:
\begin{equation}
{\boldsymbol{I}_L},{\boldsymbol{I}_H} = {\mathcal{F}_{L}}\left( {{\boldsymbol{I}}} \right),{\mathcal{F}_{H}}\left( {{\boldsymbol{I}}} \right),
\end{equation}
\begin{equation}
{\boldsymbol{I}_{LL}},{\boldsymbol{I}_{LH}},{\boldsymbol{I}_{HL}},{\boldsymbol{I}_{HH}} \!=\! {\mathcal{F}_{L}}\!\left( {{\boldsymbol{I}_L}}, {{\boldsymbol{I}_H}} \right),{\mathcal{F}_{H}}\!\left( {{\boldsymbol{I}_L}}, {{\boldsymbol{I}_H}} \right),
\end{equation}
as illustrated in the inverse wavelet transform (IDWT) shown in Fig.~\ref{fig:Transformer} (b), \hbox{$\{ {\boldsymbol{I}_{LL}},{\boldsymbol{I}_{LH}},{\boldsymbol{I}_{HL}},{\boldsymbol{I}_{HH}}\} $} can be losslessly restored to \hbox{${\boldsymbol{I}}$}. 
In contrast to the Fourier domain, the wavelet domain emphasizes high-frequency features. Therefore, performing frequency reconstruction of LR images lacking high-frequency features in the wavelet domain is a better solution.

\begin{table*}[t!]
\tiny
\setlength\tabcolsep{5pt}
\centering
\vspace{0mm}
\caption{Quantitative evaluation of our method and existing lightweight SR methods. FLOPs is measured on upsampled images with a spatial size of 1280$\times$720 pixels. The best and second-best results are emphasized in \textbf{bold} and \underline{underlined}.}
\vspace{-0mm}
\label{tab:performance}
\resizebox{0.98\textwidth}{!}{
\begin{tabular}{c|l|l|l||c|c|c|c|c}
\toprule
\rowcolor{lightgray}
& & & & \multicolumn{1}{c|}{Set5~\cite{bevilacqua2012low}} 
& \multicolumn{1}{c|}{Set14~\cite{zeyde2012single}} 
& \multicolumn{1}{c|}{BSDS100~\cite{martin2001database}} 
& \multicolumn{1}{c|}{Urban100~\cite{huang2015single}} 
& \multicolumn{1}{c}{Manga109~\cite{matsui2017sketch}} 
\\ 
\cmidrule{5-9}
    \rowcolor{lightgray}
    \multicolumn{1}{c|}{\multirow{-2}{*}{Scale}}
    & \multicolumn{1}{l|}{\multirow{-2}{*}{Methods}}  
    & \multicolumn{1}{c|}{\multirow{-2}{*}{Params}}
    & \multicolumn{1}{c||}{\multirow{-2}{*}{FLOPs}}
    & PSNR/SSIM
    & PSNR/SSIM
    & PSNR/SSIM
    & PSNR/SSIM  
    & PSNR/SSIM
    \\ 
    \hline
    \multirow{12}{*}{$\times 2$}  & CARN \cite{ahn2018fast}                           & 1592K  &222.8G   & 37.76/0.9590       & 33.52/0.9166           & 32.09/0.8978    & 31.92/0.9256     & 38.36/0.9765 \\
    & 	 HPUN-L \cite{sun2023hybrid}      		& 714K	& 151.1G	  & 38.09/0.9608       & 33.79/0.9198           & 32.25/0.9006   & 32.37/0.9307   & 39.07/0.9779 \\
    & 	DDistill-SR \cite{wang2023ddistill}     		 & 657K	& 126.4G	  & 38.08/0.9608       & 33.73/0.9195       & 32.25/0.9007   & 32.39/0.9301   & 39.16/0.9781 \\
    & 	GASSL-B \cite{wang2023global}      		 & 689K	& 158.2G	  & 38.08/0.9607       & 33.75/0.9194           & 32.24/0.9005   & 32.29/0.9298   & 38.92/0.9777 \\
    & 	SwinIR-light \cite{liang2021swinir}     		 & 878K	& 195.6G	  & 38.14/0.9611       & 33.86/0.9206           & \textbf{32.31}/\underline{0.9012}   & 32.76/\underline{0.9340}   & 39.12/\textbf{0.9783} \\
    & 	ELAN-light \cite{zhang2022efficient}      		 & \underline{582K}	& 168.4G	  & \underline{38.17}/0.9611       & \underline{33.94}/\underline{0.9207}           & \underline{32.30}/\underline{0.9012}   & 32.76/\underline{0.9340}   & 39.11/0.9782 \\
    & 	NGswin \cite{choi2023n}     		 & 998K	& 140.4G	  & 38.05/0.9610       & 33.79/0.9199       & 32.27/0.9008   & 32.53/0.9324   & 38.97/0.9777 \\
    & 	CFIN \cite{li2023cross}     		 & 675K	& \underline{116.9G}	  & 38.14/0.9613       & 33.80/0.9199       & 32.26/0.9006   & 32.48/0.9311   & 38.93/0.9347 \\
    & 	CAMixSR \cite{wang2024camixersr}     		 & 746K	& 140.6G	  & 38.16/0.9610       & 33.90/0.9206       & \textbf{32.31}/0.9010   & \underline{32.78}/0.9329   & \underline{39.25}/0.9779 \\
    & FIWHN \cite{li2022efficient}                                  &705K     &137.7G  & 38.16/\underline{0.9613}   & 33.73/0.9194        & 32.27/0.9007    &32.75/0.9337   &39.07/\underline{0.9782}     \\
    &     \textbf{DMNet (Ours)}              &\textbf{572K}     &\textbf{115.3G}  & \textbf{38.23}/\textbf{0.9613}   & \textbf{33.95}/\textbf{0.9209}        & \textbf{32.31}/\textbf{0.9015}    &\textbf{32.84}/\textbf{0.9347}   &\textbf{39.39}/0.9766     \\

    \hline

    \multirow{11}{*}{$\times 3$} &  CARN~\cite{ahn2018fast}               & 1592K   &118.8G & 34.29/0.9255        &30.29/0.8407            & 29.06/0.8034    & 28.06/0.8493   & 33.43/0.9427    \\
    &        HPUN-L~\cite{sun2023hybrid}                & 723K    &69.3G  &34.56/0.9281         &30.45/0.8445         &29.18/0.8072   &28.37/0.8572     & 33.90/0.9463 \\
    & 	 GASSL-B~\cite{wang2023global}       		& 691K	& 70.4G	  &34.47/0.9278         &30.39/0.8430          &29.15/0.8063   &28.27/0.8546     & 33.77/0.9455 \\
    & 	DDistill-SR \cite{wang2023ddistill}      		 & 665K	& 56.9G	  & 34.43/0.9276       & 30.39/0.8432         & 29.16/0.8070   & 28.31/0.8546   & 33.97/0.9465 \\
    & 	SwinIR-light \cite{liang2021swinir}     	 & 886K	& 87.2G	  & 34.62/\underline{0.9289}       & 30.54/\underline{0.8463}           & 29.20/0.8082   & 28.66/\underline{0.8624}   & 33.98/\underline{0.9478} \\
    & 	ELAN-light \cite{zhang2022efficient}     		 & \underline{590K}	& 75.7G	  & 34.61/0.9288       & \underline{30.55}/\textbf{0.8463}           & \underline{29.21}/0.8081   & \underline{28.69}/\underline{0.8624}   & \underline{34.00}/\underline{0.9478} \\
    & 	NGswin \cite{choi2023n}      		 & 1007K	& 66.6G	  & 34.52/0.9282       & 30.53/0.8456         & 29.19/0.8078   & 28.52/0.8603   & 33.89/0.9470 \\
    & 	CFIN \cite{li2023cross}      		 & 681K	& \underline{53.5G}	  & 34.65/0.9289       & 30.45/0.8443         & 29.18/0.8071   & 28.49/0.8583   & 33.89/0.9464 \\
    & FIWHN \cite{li2022efficient}                                     &713K     &62.0G  &  34.50/0.9283  & 30.50/0.8451   & 29.19/0.8077    & 28.62/0.8607    &33.97/0.9472  \\
    &  \textbf{DMNet (Ours)}            &\textbf{579K}     &\textbf{52.0G}  &  \textbf{34.71}/\textbf{0.9295}  & \textbf{30.57}/0.8459   & \textbf{29.26}/\textbf{0.8093}    & \textbf{28.80}/\textbf{0.8640}    & \textbf{34.33}/\textbf{0.9488}  \\

    \hline

    \multirow{12}{*}{$\times 4$} &   CARN~\cite{ahn2018fast}              & 1592K  & 90.9G & 32.13/0.8937        & 28.60/0.7806            & 27.58/0.7349    & 26.07/0.7837   & 30.42/0.9070    \\
    & HPUN-L~\cite{sun2023hybrid}                   & 734K  & 39.7G   & 32.31/0.8962   &28.73/0.7842            &27.66/0.7386        &26.27/0.7918                & 30.77/0.9109 \\
    & GASSL-B~\cite{wang2023global}       	 & 694K  & 39.9G   & 32.27/0.8962   &28.74/0.7850            &27.66/0.7388        &26.27/0.7914                 & 30.92/0.9122 \\
    & 	DDistill-SR \cite{wang2023ddistill}      		 & 675K  & 32.6G   & 32.29/0.8961   &28.69/0.7833       &27.65/0.7385        &26.25/0.7893                 & 30.79/0.9098 \\
    & SwinIR-light~\cite{liang2021swinir}         	 & 897K  & 49.6G   &32.44/0.8976   &28.77/0.7858            &\underline{27.69}/\underline{0.7406}        &26.47/0.7980                 & 30.92/\textbf{0.9151} \\
    & ELAN-light~\cite{zhang2022efficient}        	 & \underline{601K}  & 43.2G   & 32.43/0.8975   &\underline{28.78}/0.7858            &\underline{27.69}/\underline{0.7406}        &26.54/0.7982                 & 30.92/0.9150 \\
    & NGswin~\cite{choi2023n}        	 & 1019K  & 36.4G   & 32.33/0.8963   &\underline{28.78}/\underline{0.7859}            &27.66/0.7396        &26.45/0.7963                 & 30.80/0.9128 \\
    & CFIN~\cite{li2023cross}        	 & 699K  & \underline{31.2G}   & 32.49/0.8985   &28.74/0.7849            &27.68/0.7396        &26.39/0.7946                 & 30.73/0.9124 \\
    & CAMixSR~\cite{wang2024camixersr}        	 & 765K  & 37.8G   &  \underline{32.45}/ \underline{0.8978}   &\underline{28.78}/0.7856            &\underline{27.69}/0.7401        &26.51/0.7966                 & \underline{31.06}/0.9148 \\
    & FIWHN~\cite{li2022efficient}                                  &725K  &35.6G  & 32.30/0.8967    & 28.76/0.7849    & 27.68/0.7400   & \underline{26.57}/\underline{0.7989}       &30.93/0.9131     \\
    & \textbf{DMNet (Ours)}                       &\textbf{588K}  &\textbf{29.7G}  & \textbf{32.51}/\textbf{0.8987}    & \textbf{28.84}/\textbf{0.7866}    & \textbf{27.73}/\textbf{0.7410}   &  \textbf{26.58}/\textbf{0.7991}       & \textbf{31.14}/\underline{0.9150}     \\

    \toprule
\end{tabular}}
\end{table*}

\paragraph{Wavelet-domain Modulation Self-attention (WMA)} In our WMA, we employ self-attention and dynamic convolution to modulate different frequency subbands obtained through DWT. Then, we utilize IDWT to restore reconstructed frequency-domain subbands to output features. Therefore, the output will exhibit more accurate frequency domain characteristics. Specifically, as shown in Fig.~\ref{fig:Transformer} (c), for an input feature \hbox{$\boldsymbol{F}_{in}^{W}$~$\in$~$\mathbb{R}^{H\times W \times C}$}, to reduce Params counts, we first reduce the channels of \hbox{$\boldsymbol{F}_{in}^{W}$} from \hbox{$C$} to \hbox{$C/4$} using a convolution layer. Then we utilize DWT to decompose features and concatenate the resulting four wavelet subbands along channels to form frequency feature \hbox{$ {\boldsymbol{F}_{DWT}} $~$\in$~$\mathbb{R}^{\frac{H}{2}\times \frac{W}{2} \times C}$}: 
\begin{equation}
{\boldsymbol{F}_{\scriptscriptstyle DWT}} \!=\! \left[ {{\boldsymbol{F}_{\scriptscriptstyle LL}},\!{\boldsymbol{F}_{\scriptscriptstyle LH}},\!{\boldsymbol{F}_{\scriptscriptstyle HL}},\!{\boldsymbol{F}_{\scriptscriptstyle HH}}} \right] \!=\! \mathcal{DWT}\!\left(\! {\mathcal{R}\!\left(\! {\boldsymbol{F}_{in}^{W}}\! \right)} \!\right),
\end{equation}
where \hbox{$\mathcal{R}$} is channel reduction operator, \hbox{$\mathcal{DWT}$} is discrete wavelet transform. Next, we extend the frequency domain features of the existing channels and transform to obtain vectors \hbox{$\left\{ {\boldsymbol{Q}', \boldsymbol{K}', \boldsymbol{V}'} \right\}$~$\in$~$\mathbb{R}^{C \times \frac{HW}{4}}$} containing different frequency domain components. As shown in Fig.~\ref{fig:Transformer} (c), in channel components of vectors \hbox{${\boldsymbol{Q}'}$}, \hbox{${\boldsymbol{K}'}$} and \hbox{${\boldsymbol{V}'}$}, the first component is low-frequency component \hbox{${\boldsymbol{F}_{LL}}$} and other components is high-frequency component \hbox{$\left\{ {\boldsymbol{F}_{LH}},{\boldsymbol{F}_{HL}},{\boldsymbol{F}_{HH}} \right\}$}.
\begin{equation}
\left\{ {\boldsymbol{Q}'},{\boldsymbol{K}'},{\boldsymbol{V}'} \right\} = {{\mathcal{RS}}\left( {{\mathcal{D}}\left( {{\mathcal{P}}\left( {{\boldsymbol{F}_{DWT}}} \right)} \right)} \right)},
\end{equation}
where ${\mathcal{RS}}$ denotes a reshape operation, ${\mathcal{D}}$ refers to a 3×3 depth-wise convolution, and ${\mathcal{P}}$ represents a 1×1 point-wise convolution. Then, we use self-attention to explore the frequency domain. Specifically, we compute the similarity scores between high- and low-frequency features within the \hbox{${\boldsymbol{Q}'}$} and \hbox{${\boldsymbol{K}'}$} vectors, and use these scores to adaptively extract features from the \hbox{${\boldsymbol{V}'}$} vector, enabling the self-attention to complement high- and low-frequency information from wavelet features.
\begin{equation}
{\rm{Attention}}({\boldsymbol{Q}'},{\boldsymbol{K}'},{\boldsymbol{V}'}) = {\rm{Softmax}}({\boldsymbol{Q}'}{{\boldsymbol{K}'}^T}/\alpha ){\boldsymbol{V}'},
\end{equation}
since self-attention does not capture detailed features effectively, we introduce dynamic convolution to adaptively aid our WMA in refining frequency details. Initially, a large kernel group convolution \hbox{$\mathcal{G}$} is applied to produce a set of dynamic weights \hbox{$\boldsymbol{W}$} from wavelet features \hbox{$\boldsymbol{F}_{DWT}$}. Then, \hbox{$\boldsymbol{W}$} is used to adaptively modulate features \hbox{$\boldsymbol{F}_{Self}$} obtained from self-attention, resulting in refined features \hbox{${\boldsymbol{F}_{SD}}$~$\in$~$\mathbb{R}^{\frac{H}{2}\times \frac{W}{2} \times C}$}:
\begin{equation}
{\boldsymbol{F}_{SD}} = {\rm{Attention}}\left( {{\boldsymbol{Q}'}}, {{\boldsymbol{K}'}}, {{\boldsymbol{V}'}} \right) \circledast \mathcal{G}\left( {{\boldsymbol{F}_{DWT}}} \right),
\end{equation}
where $\circledast$ is the dynamic convolution. Finally, \hbox{${\boldsymbol{F}_{SD}}$} is reduced to a spatial domain feature after inverse discrete wavelet transform \hbox{$\mathcal{IDWT}$}. Since its channel count is only \hbox{$C/4$}, we need perform a channel expansion \hbox{$\mathcal{E}$} to obtain \hbox{$\boldsymbol{F}_{out}^{W}$~$\in$~$\mathbb{R}^{H\times W \times C}$}:
\begin{equation}
\boldsymbol{F}_{out}^{W} = \mathcal{E}\left( {\mathcal{IDWT}\left( {{\boldsymbol{F}_{SD}}} \right)} \right),
\end{equation}
after the modulation process described above, our WMA can accurately recover frequency-domain characteristics, particularly high-frequency features crucial for reconstruction.

\subsection{Spatial-domain Modulation Transformer}
To reconstruct spatial features, as shown in Fig.~\ref{fig:main}, we propose an SMT. Specifically, we first utilize spatial domain modulation self-attention to estimate spatial features. Subsequently, a feed-forward network~\cite{zamir2022restormer} ($\mathcal{FFN}$) is utilized to aggregate features to refine representations. For an input \hbox{$\boldsymbol{X}_{in}^{S}$~$\in$~$\mathbb{R}^{H\times W \times C}$}, the process of obtaining output \hbox{${\boldsymbol{X}_{out}^{S}}$} is:
\begin{equation}
\hat{\boldsymbol{X}}^{S}   = \mathcal{SMA}\left( {\mathcal{LN}\left( {{\boldsymbol{X}_{in}^{S}}} \right)} \right) + \boldsymbol{X}_{in}^{S} ,
\end{equation}
\begin{equation}
{\boldsymbol{X}_{out}^{S}}   = \mathcal{FFN}\left( {\mathcal{LN}\left( \hat{\boldsymbol{X}}^{S} \right)} \right) + \hat{\boldsymbol{X}}^{S} ,
\end{equation}
where $\mathcal{LN}$ is layer normalization and $\mathcal{SMA}$ is spatial-domain modulation self-attention. The core part is spatial-domain modulation self-attention; we will detail this component below.

\paragraph{Spatial-domain Modulation Self-attention (SMA)} In our SMA, to estimate spatial-domain features with less consumption, our SMA condenses extracted spatial domain features in channels and implicitly estimates global spatial domain features by performing self-attention over channels.

\begin{table*}[t!]
\tiny
\setlength\tabcolsep{2pt}
\centering
\vspace{0mm}
\caption{Quantitative evaluation with Transformer/Mamba-based lightweight SR methods. FLOPs and speed are measured on upsampled images
with a spatial size of 1280×720 pixels. Memory is measured when batch size is 16 and patch size is 64.}
\vspace{-0mm}
\label{tab:trans_mamba_performance}
\resizebox{0.98\textwidth}{!}{
\begin{tabular}{c|l|l|l|l|l||c|c|c|c|c|c}
\toprule
\rowcolor{lightgray}
& & & & & & \multicolumn{1}{c|}{Set5~\cite{bevilacqua2012low}} 
& \multicolumn{1}{c|}{Set14~\cite{zeyde2012single}} 
& \multicolumn{1}{c|}{BSDS100~\cite{martin2001database}} 
& \multicolumn{1}{c|}{Urban100~\cite{huang2015single}} 
& \multicolumn{1}{c|}{Manga109~\cite{matsui2017sketch}} 
& \multicolumn{1}{c}{Average} 
\\ 
\cmidrule{7-12}
    \rowcolor{lightgray}
    \multicolumn{1}{c|}{\multirow{-2}{*}{Scale}}
    & \multicolumn{1}{l|}{\multirow{-2}{*}{Methods}}  
    & \multicolumn{1}{c|}{\multirow{-2}{*}{Params}}
    & \multicolumn{1}{c|}{\multirow{-2}{*}{FLOPs}}
    & \multicolumn{1}{c|}{\multirow{-2}{*}{Speed}}
    & \multicolumn{1}{c||}{\multirow{-2}{*}{Memory}}
    & PSNR/SSIM
    & PSNR/SSIM
    & PSNR/SSIM
    & PSNR/SSIM  
    & PSNR/SSIM
    & PSNR/SSIM
    \\ 
    \hline
    \multirow{7}{*}{$\times 2$} 
    & 	SwinIR-light \cite{liang2021swinir}     		 & 878K	& 195.6G   & 1.30s & \underline{8.0G} & 38.14/0.9611       & 33.86/0.9206           & 32.31/0.9012   & 32.76/0.9340   & 39.12/0.9783 & 35.24/0.9390 \\
    & 	DLGSANet-light \cite{zhou2023srformer}     		 & 745K	& 170.0G	& 0.11s & 9.9G & 38.20/0.9612       & 33.89/0.9203       & 32.30/0.9012   & 32.94/0.9355   & 39.29/0.9780 & 35.32/0.9392\\
    & 	SRFormer-light \cite{zhou2023srformer}     		 & 853K	& 236.2G	& 1.39s & 8.7G & \underline{38.23}/\underline{0.9613}       & 33.94/0.9209       & 32.36/0.9019   & 32.91/0.9353   & 39.28/0.9779 & 35.36/0.9394\\
    & 	HiT-SNG \cite{zhang2025hit}     		 & 1013K	& 213.9G	& 0.24s & 16.6G & 38.21/0.9612       & \textbf{34.00}/\textbf{0.9217}       & \textbf{32.35}/\textbf{0.9019}   & \textbf{33.01}/\textbf{0.9360}   & \underline{39.32}/\underline{0.9782} & \underline{35.38}/\underline{0.9396} \\
    & 	MambaIR \cite{guo2025mambair}     		 & 879K	& 198.1G	& 0.49s & 16.7G & 38.16/0.9610       & \textbf{34.00}/\underline{0.9212}       & \underline{32.34}/\underline{0.9017}   & 32.92/0.9356   & 39.31/0.9779 & 35.35/0.9394\\
    
    &     \textbf{DMNet (Ours)}              &\textbf{572K}     &\textbf{115.3G}  & \textbf{0.07s} & \textbf{7.0G} & \underline{38.23}/\underline{0.9613}   & \underline{33.95}/0.9209        & 32.31/0.9015    &32.84/0.9347   &\underline{39.39}/0.9766    & 35.35/0.9391 \\
    &     \textbf{DMNet-L (Ours)}              &\underline{733K}     &\underline{147.4G}  & \underline{0.09s} & 9.1G & \textbf{38.27}/\textbf{0.9615}   & \textbf{34.00}/\underline{0.9212}        & \underline{32.34}/\underline{0.9017}    &\underline{32.98}/\underline{0.9359}   &\textbf{39.52}/\textbf{0.9791}   &\textbf{35.42}/\textbf{0.9399} \\

    \toprule
\end{tabular}}
\end{table*}

As shown in Fig.~\ref{fig:Transformer} (a), for an input \hbox{$\boldsymbol{F}_{in}^{S}$~$\in$~$\mathbb{R}^{H\times W \times C}$}, we first utilize a \hbox{$1 \times 1$} point-wise convolution \hbox{${\mathcal{P}}$} and a \hbox{$3 \times 3$} depth-wise convolution \hbox{${\mathcal{D}}$} to expand channels and extract features. Then, query \hbox{$\boldsymbol{Q}$~$\in$~$\mathbb{R}^{C \times HW}$}, key \hbox{$\boldsymbol{K}$~$\in$~$\mathbb{R}^{C \times HW}$}, and value \hbox{$\boldsymbol{V}$~$\in$~$\mathbb{R}^{C \times HW}$} are obtained by performing a reshape operation \hbox{${\mathcal{RS}}$} on channel space. This process is:
\begin{equation}
\left\{ {\boldsymbol{Q}},{\boldsymbol{K}},{\boldsymbol{V}} \right\} = {{\mathcal{RS}}\left( {{\mathcal{D}}\left( {{\mathcal{P}}\left( {{\boldsymbol{F}_{in}^{S}}} \right)} \right)} \right)},
\end{equation}
next, normalization-operated vectors \hbox{$\boldsymbol{Q}$}, \hbox{$\boldsymbol{K}$}, \hbox{$\boldsymbol{V}$} explore potential relationships of spatial domain features within channels through self-attention, which in turn reconstruct spatial domain pixel distribution. This process can be formulated as:
\begin{equation}
{\rm{Attention}}({\boldsymbol{Q}},{\boldsymbol{K}},{\boldsymbol{V}}) = {\rm{Softmax}}({\boldsymbol{Q}}{{\boldsymbol{K}}^T}/\alpha ){\boldsymbol{V}},
\end{equation}
where $\alpha$ is a learnable factor, after several SMA and WMA blocks, we get a depth feature \hbox{${{\boldsymbol{F}_{d}}}$}. Finally, we can get an SR image \hbox{${{\boldsymbol{I}_{SR}}}$} from \hbox{${{\boldsymbol{F}_{d}}}$} after a Pixel-Shuffle.

\subsection{Loss Function}~\label{Loss}
For image pairs \hbox{$\left\{ {\boldsymbol{I}_{SR},\boldsymbol{I}_{HR}} \right\}$} in training, a reconstruction loss \hbox{${\mathcal{L}_{pixel}}$} is used for supervising spatial-domain SR:
\begin{equation}
{\mathcal{L}_{pixel}} = {{{\left\| {\boldsymbol{I}_{SR} - \boldsymbol{I}_{HR}} \right\|}_1}},
\end{equation}
where \hbox{${\boldsymbol{I}_{SR}}$} is the SR result of our method, \hbox{${\boldsymbol{I}_{HR}}$} is ground truth. Further, we introduce the frequency loss \hbox{${\mathcal{L}}_{fre}$} in our frequency domain output \hbox{${\boldsymbol{I}_{Fre}}$} of amplitude \hbox{${\boldsymbol{A}_{SR}}$} and phase \hbox{${\boldsymbol{P}_{SR}}$} to ensure the accuracy of the frequency domain SR: 
\begin{equation}
{\mathcal{L}_{fre}} = {{{\left\| {\left[ {\boldsymbol{A}_{SR},\boldsymbol{P}_{SR}} \right] - \left[ {\boldsymbol{A}_{HR},\boldsymbol{P}_{HR}} \right]} \right\|}_1}},
\end{equation}
where \hbox{${\boldsymbol{A}_{HR}}$} and \hbox{${\boldsymbol{P}_{HR}}$} are the amplitude and phase of frequency domain ground truth \hbox{${\boldsymbol{G}_{Fre}}$}. The Fourier loss promotes model convergence more effectively than the potential conflicts between different wavelet subbands during gradient updates, which require careful parameter tuning. Therefore, We use Fourier loss. Finally, our total loss is:
\begin{equation}
{\mathcal{L}_{total}} = {\mathcal{L}_{pixel}} + \lambda {\mathcal{L}_{fre}}, \quad \lambda = {10^{ - 1}}.
\label{eq: 16}
\end{equation}

\section{Experiments}
\subsection{Experimental Settings}
\subsubsection{Datasets} Our method is trained on the DIV2K~\cite{timofte2017ntire} dataset which contains 800 pairs of training images. To evaluate our method, we test on Set5 ~\cite{bevilacqua2012low}, Set14~\cite{zeyde2012single}, BSDS100~\cite{martin2001database}, Urban100~\cite{huang2015single}, and Manga109~\cite{matsui2017sketch}. We also train and evaluate our method on a real-world dataset RealSRv3~\cite{cai2019toward}, which obtains LR-HR image pairs on the same scene by adjusting the focal length of a digital camera and alignment operations to images to simulink blurred LR images.

\begin{table}[t!]
\tiny
\setlength\tabcolsep{2.5pt}
\centering
\vspace{0mm}
\caption{Quantitative evaluation of our method and existing lightweight SR methods on a real-world test set.}
\vspace{-0mm}
\label{tab:realsr}
\resizebox{0.47\textwidth}{!}{
\begin{tabular}{c|l|l|l||c}
\toprule
\rowcolor{lightgray}
& & & & \multicolumn{1}{c}{RealSRv3~\cite{cai2019toward}} 
\\ 
\cmidrule{5-5}
    \rowcolor{lightgray}
    \multicolumn{1}{c|}{\multirow{-2}{*}{Scale}}
    & \multicolumn{1}{l|}{\multirow{-2}{*}{Methods}} 
    & \multicolumn{1}{c|}{\multirow{-2}{*}{Params}}
    & \multicolumn{1}{c||}{\multirow{-2}{*}{FLOPs}}
    & PSNR$\uparrow$/SSIM$\uparrow$/LPIPS$\downarrow$
    \\ 
    \hline
    \multirow{6}{*}{$\times 4$}  & IMDN~\cite{hui2019lightweight}     & 715K  &40.9G   & 29.20/0.8264/0.2810     \\
    & SwinIR-light~\cite{liang2021swinir}    & 897K  &49.6G   & 29.18/0.8278/0.2756     \\
    & ELAN-light~\cite{zhang2022efficient}  & 601K  &43.2G   & 29.20/0.8272/0.2764     \\
    & NGswin~\cite{choi2023n}  & 1019K  &36.4G   & 29.06/0.8240/0.2779     \\
    & MambaIR~\cite{guo2025mambair}  & 879K  &50.6G   & 29.22/0.8286/0.2748     \\
    & \textbf{DMNet (Ours)}   & \textbf{588K}  & \textbf{29.7G}   & \textbf{29.27}/\textbf{0.8295}/\textbf{0.2739}    \\

    \bottomrule
\end{tabular}}
\end{table}

\subsubsection{Metrics and Implementations} We calculate PSNR and SSIM~\cite{wang2002universal} on the Y channel in the YCbCr color space. We additionally employ LPIPS~\cite{zhang2018unreasonable} for evaluating real test sets. During training, we set the batch size to 64, the initial learning rate to \hbox{$5 \times {10^{ - 4}}$} with a scheduler in 500K iterations, and the patch size is set to \hbox{$64 \times 64$}. We train DMNet using Adam optimizer with \hbox{${\beta _{\rm{1}}}$}=0.9, \hbox{${\beta _{\rm{2}}}$}=0.99. We implement all experiments using the Pytorch framework with the NVIDIA GeForce RTX 4090. In our DMNet, we use the test-time local converter~\cite{chu2022improving} during testing. we set \hbox{${N_1}$} to 3, \hbox{${N_2}$} to 3, and the number of channels to 48. Inference time is measured at input resolution $(1280/s, 720/s)$, where $s$ is the upscaling factor. Results are averaged over 100 runs with batch size of 1 for consistency.

\subsection{Comparisons with State-of-the-Art Methods}
We compare our method with several state-of-the-art methods, including CNN-based methods CARN~\cite{ahn2018fast}, IMDN~\cite{hui2019lightweight}, GASSL-B~\cite{wang2023global} and Transformer-based methods SwinIR-light~\cite{liang2021swinir}, ELAN-light~\cite{zhang2022efficient}, NGswin~\cite{choi2023n}, CAMixSR~\cite{wang2024camixersr}.

   


\begin{figure*}[htpb]
	\scriptsize
	\centering
	\scalebox{0.9}{
		\begin{tabular}{lc}
            \begin{adjustbox}{valign=t}
				\begin{tabular}{c}				    
                    \includegraphics[width=0.3\textwidth, height=0.145\textheight]{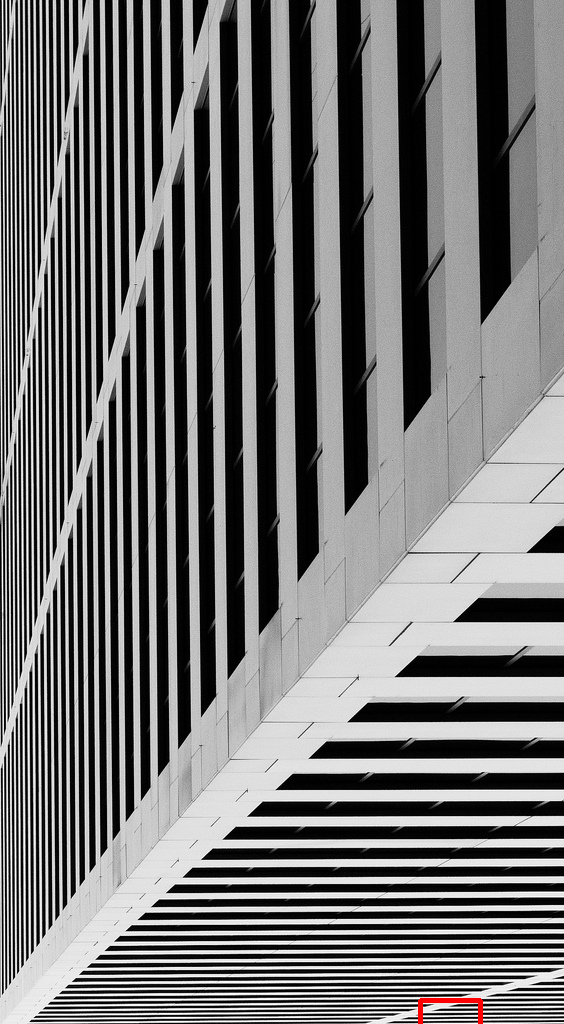} \\
					\fontsize{7.5pt}{1pt}\selectfont Urban100 ($\times 2$): img011 \\
				\end{tabular}
			\end{adjustbox}
			\hspace{-3mm}
			\begin{adjustbox}{valign=t}
				\begin{tabular}{ccccc}
                    \includegraphics[width=0.15\textwidth, height=0.065\textheight]{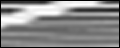} & 
					\hspace{-3mm}
					\includegraphics[width=0.15\textwidth, height=0.065\textheight]{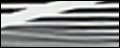} & 
					\hspace{-3mm}
					\includegraphics[width=0.15\textwidth, height=0.065\textheight]{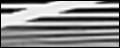} & 
                    \hspace{-3mm}
                    \includegraphics[width=0.15\textwidth, height=0.065\textheight]{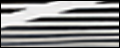} & 
                    \hspace{-3mm}
                    \includegraphics[width=0.15\textwidth, height=0.065\textheight]{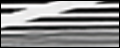}\\
					\fontsize{7.5pt}{1pt}\selectfont LR & \hspace{-2mm}
				  \fontsize{7.5pt}{1pt}\selectfont 
                    CARN~\cite{ahn2018fast} & \hspace{-2mm}
					\fontsize{7.5pt}{1pt}\selectfont IMDN~\cite{hui2019lightweight} & \hspace{-2mm}
                    \fontsize{7.5pt}{1pt}\selectfont FDIWN~\cite{gao2022feature} & \hspace{-2mm}
                    \fontsize{7.5pt}{1pt}\selectfont LBNet~\cite{gao2022lightweight}\\
                    \includegraphics[width=0.15\textwidth, height=0.065\textheight]{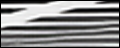} & 
                    \hspace{-3mm}
					\includegraphics[width=0.15\textwidth, height=0.065\textheight]{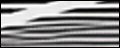} & 
					\hspace{-3mm}
					\includegraphics[width=0.15\textwidth, height=0.065\textheight]{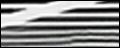} & 
					\hspace{-3mm}
                    \includegraphics[width=0.15\textwidth, height=0.065\textheight]{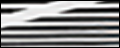} & 
					\hspace{-3mm}
					\includegraphics[width=0.15\textwidth, height=0.065\textheight]{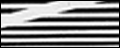} \\
                    \fontsize{7.5pt}{1pt}\selectfont SwinIR-light~\cite{liang2021swinir} & \hspace{-2mm}
					\fontsize{7.5pt}{1pt}\selectfont ELAN-light~\cite{zhang2022efficient} & \hspace{-2mm}
				  \fontsize{7.5pt}{1pt}\selectfont 
                    NGSwin~\cite{choi2023n} & 
                    \hspace{-2mm}
                    \fontsize{7.5pt}{1pt}\selectfont \textbf{Ours} & 
                    \hspace{-2mm}
					\fontsize{7.5pt}{1pt}\selectfont Grouth Truth \\
				\end{tabular}
			\end{adjustbox}
             \\

             \begin{adjustbox}{valign=t}
				\begin{tabular}{c}
                \includegraphics[width=0.3\textwidth, height=0.145\textheight]{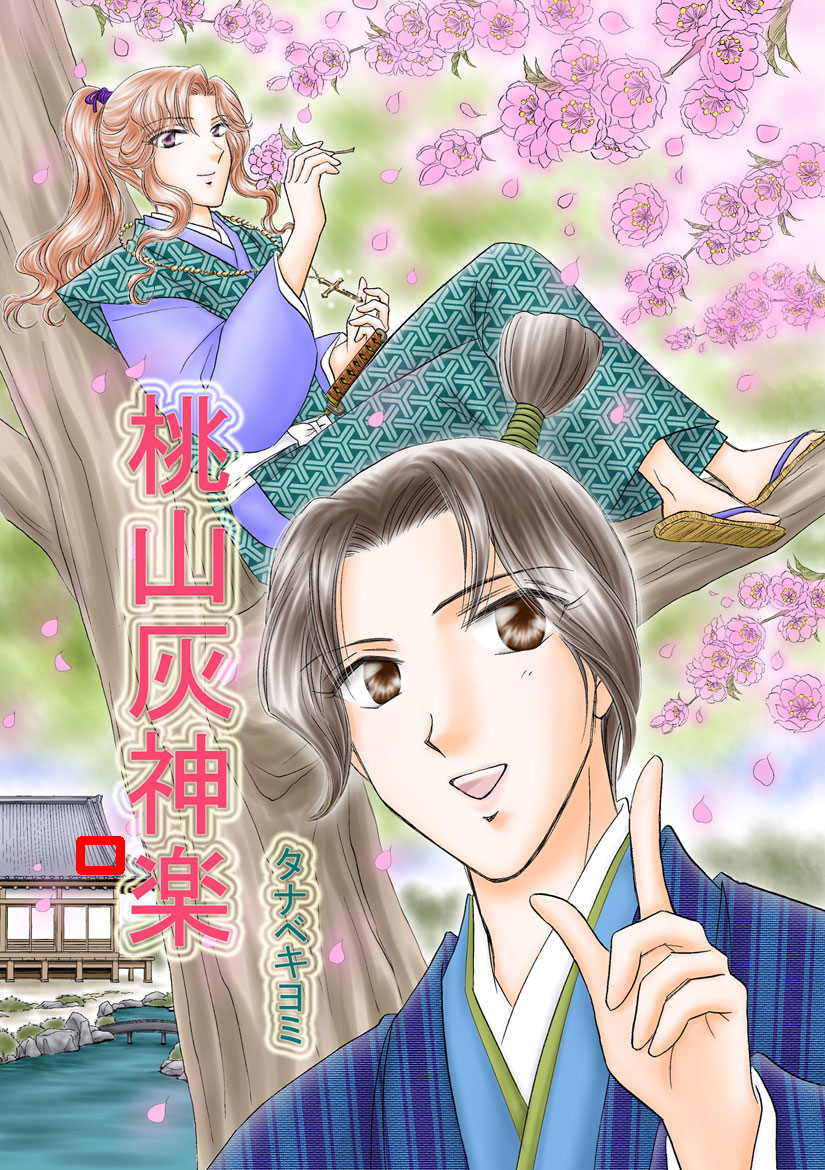} \\
					\fontsize{7.5pt}{1pt}\selectfont Manga109 ($\times 3$): MomoyamaHaikagura \\
				\end{tabular}
			\end{adjustbox}
			\hspace{-3mm}
			\begin{adjustbox}{valign=t}
				\begin{tabular}{ccccc}
                    \includegraphics[width=0.15\textwidth, height=0.065\textheight]{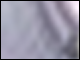} & 
					\hspace{-3mm}
					\includegraphics[width=0.15\textwidth, height=0.065\textheight]{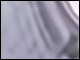} & 
					\hspace{-3mm}
					\includegraphics[width=0.15\textwidth, height=0.065\textheight]{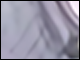} & 
                    \hspace{-3mm}
                    \includegraphics[width=0.15\textwidth, height=0.065\textheight]{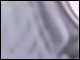} & 
                    \hspace{-3mm}
                    \includegraphics[width=0.15\textwidth, height=0.065\textheight]{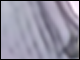}\\
					\fontsize{7.5pt}{1pt}\selectfont LR & \hspace{-2mm}
				  \fontsize{7.5pt}{1pt}\selectfont 
                    CARN~\cite{ahn2018fast} & \hspace{-2mm}
					\fontsize{7.5pt}{1pt}\selectfont IMDN~\cite{hui2019lightweight} & \hspace{-2mm}
                    \fontsize{7.5pt}{1pt}\selectfont FDIWN~\cite{gao2022feature} & \hspace{-2mm}
                    \fontsize{7.5pt}{1pt}\selectfont LBNet~\cite{gao2022lightweight}\\
                    \includegraphics[width=0.15\textwidth, height=0.065\textheight]{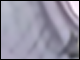} & 
                    \hspace{-3mm}
					\includegraphics[width=0.15\textwidth, height=0.065\textheight]{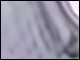} & 
					\hspace{-3mm}
					\includegraphics[width=0.15\textwidth, height=0.065\textheight]{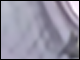} & 
					\hspace{-3mm}
                    \includegraphics[width=0.15\textwidth, height=0.065\textheight]{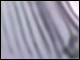} & 
					\hspace{-3mm}
					\includegraphics[width=0.15\textwidth, height=0.065\textheight]{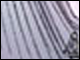} \\
                    \fontsize{7.5pt}{1pt}\selectfont SwinIR-light~\cite{liang2021swinir} & \hspace{-2mm}
					\fontsize{7.5pt}{1pt}\selectfont ELAN-light~\cite{zhang2022efficient} & \hspace{-2mm}
				  \fontsize{7.5pt}{1pt}\selectfont 
                    NGSwin~\cite{choi2023n} & 
                    \hspace{-2mm}
                    \fontsize{7.5pt}{1pt}\selectfont \textbf{Ours} & 
                    \hspace{-2mm}
					\fontsize{7.5pt}{1pt}\selectfont Grouth Truth \\
				\end{tabular}
			\end{adjustbox}
             \\

             \begin{adjustbox}{valign=t}
				\begin{tabular}{c}
					\includegraphics[width=0.3\textwidth, height=0.145\textheight]{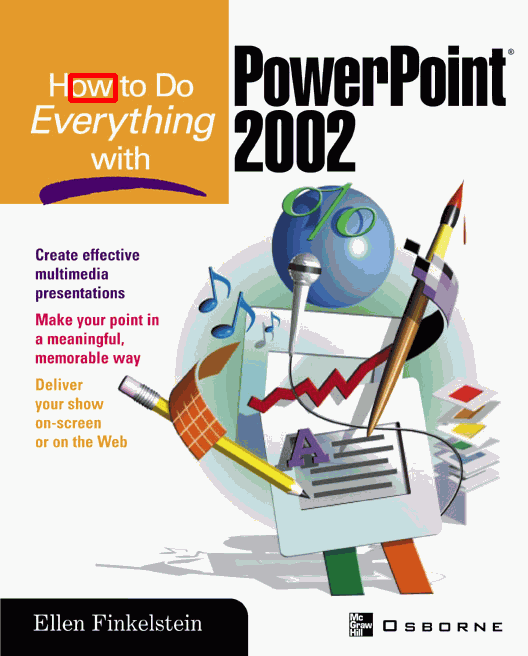} \\
					\fontsize{7.5pt}{1pt}\selectfont Set14 ($\times 4$): ppt \\
				\end{tabular}
			\end{adjustbox}
			\hspace{-3mm}
			\begin{adjustbox}{valign=t}
				\begin{tabular}{ccccc}
					\includegraphics[width=0.15\textwidth, height=0.065\textheight]{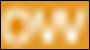} & 
					\hspace{-3mm}
					\includegraphics[width=0.15\textwidth, height=0.065\textheight]{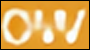} & 
					\hspace{-3mm}
					\includegraphics[width=0.15\textwidth, height=0.065\textheight]{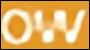} & 
                    \hspace{-3mm}
                    \includegraphics[width=0.15\textwidth, height=0.065\textheight]{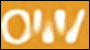} & 
                    \hspace{-3mm}
                    \includegraphics[width=0.15\textwidth, height=0.065\textheight]{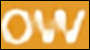}\\
					\fontsize{7.5pt}{1pt}\selectfont LR & \hspace{-2mm}
				  \fontsize{7.5pt}{1pt}\selectfont CARN~\cite{ahn2018fast} & \hspace{-2mm}
					\fontsize{7.5pt}{1pt}\selectfont IMDN~\cite{hui2019lightweight} & \hspace{-2mm}
                    \fontsize{7.5pt}{1pt}\selectfont FDIWN~\cite{gao2022feature} & \hspace{-2mm}
                    \fontsize{7.5pt}{1pt}\selectfont LBNet~\cite{gao2022lightweight}\\
                    \includegraphics[width=0.15\textwidth, height=0.065\textheight]{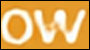} & 
                    \hspace{-3mm}
					\includegraphics[width=0.15\textwidth, height=0.065\textheight]{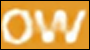} & 
					\hspace{-3mm}
					\includegraphics[width=0.15\textwidth, height=0.065\textheight]{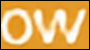} & 
					\hspace{-3mm}
                    \includegraphics[width=0.15\textwidth, height=0.065\textheight]{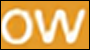} & 
					\hspace{-3mm}
					\includegraphics[width=0.15\textwidth, height=0.065\textheight]{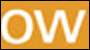} \\
                    \fontsize{7.5pt}{1pt}\selectfont SwinIR-light~\cite{liang2021swinir} & \hspace{-2mm}
					\fontsize{7.5pt}{1pt}\selectfont ELAN-light~\cite{zhang2022efficient} & \hspace{-2mm}
				  \fontsize{7.5pt}{1pt}\selectfont NGSwin~\cite{choi2023n} & 
                    \hspace{-2mm}
                    \fontsize{7.5pt}{1pt}\selectfont \textbf{Ours} & 
                    \hspace{-2mm}
					\fontsize{7.5pt}{1pt}\selectfont Grouth Truth \\
				\end{tabular}
			\end{adjustbox}
             \\
			
	\end{tabular} }
	\vspace{-0mm}
	\caption{Qualitative comparisons with existing methods on the synthesized test set. Our method can reconstruct clearer edges.}
	\label{visual}
\end{figure*}

\begin{figure*}[!t]
	\scriptsize
	\centering
	\scalebox{0.9}{
		\begin{tabular}{lc}
            \begin{adjustbox}{valign=t}
				\begin{tabular}{c}
				\includegraphics[width=0.23\textwidth, height=0.125\textheight]{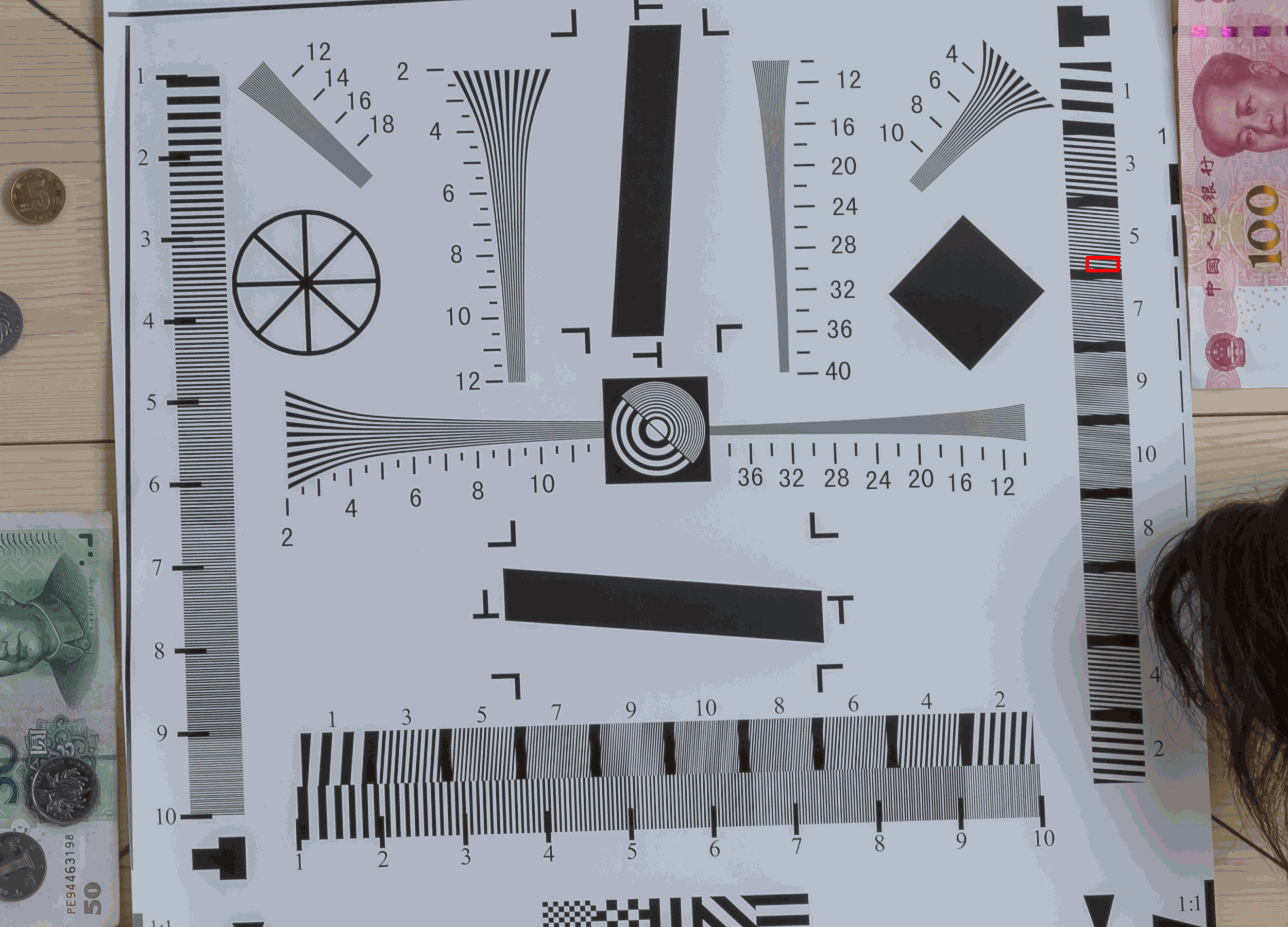} \\
					\fontsize{6.5pt}{1pt}\selectfont RealSRv3 ($\times 4$): Canon045 \\
				\end{tabular}
			\end{adjustbox}
			\hspace{-3mm}
			\begin{adjustbox}{valign=t}
				\begin{tabular}{ccc}
					\includegraphics[width=0.09\textwidth, height=0.055\textheight]{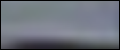} & 
					\hspace{-3mm}
                    \includegraphics[width=0.09\textwidth, height=0.055\textheight]{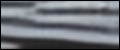} & 
					\hspace{-3mm}
					\includegraphics[width=0.09\textwidth, height=0.055\textheight]{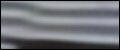} \\
					\fontsize{6.5pt}{1pt}\selectfont LR & \hspace{-3mm}
					\fontsize{6.5pt}{1pt}\selectfont SwinIR-light~\cite{liang2021swinir} & \hspace{-3mm}
                    \fontsize{6.5pt}{1pt}\selectfont ELAN-light~\cite{zhang2022efficient}\\
					\includegraphics[width=0.09\textwidth, height=0.055\textheight]{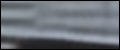} & 
					\hspace{-3mm}

                    \includegraphics[width=0.09\textwidth, height=0.055\textheight]{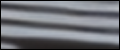} & 
					\hspace{-3mm}
					\includegraphics[width=0.09\textwidth, height=0.055\textheight]{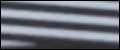} \\
					\fontsize{6.5pt}{1pt}\selectfont NGswin~\cite{choi2023n} & \hspace{-3mm}
                    \fontsize{6.5pt}{1pt}\selectfont \textbf{Ours} & \hspace{-3mm}
					\fontsize{6.5pt}{1pt}\selectfont Ground Truth \\
				\end{tabular}
			\end{adjustbox}

            \begin{adjustbox}{valign=t}
				\begin{tabular}{c}
					\includegraphics[width=0.23\textwidth, height=0.125\textheight]{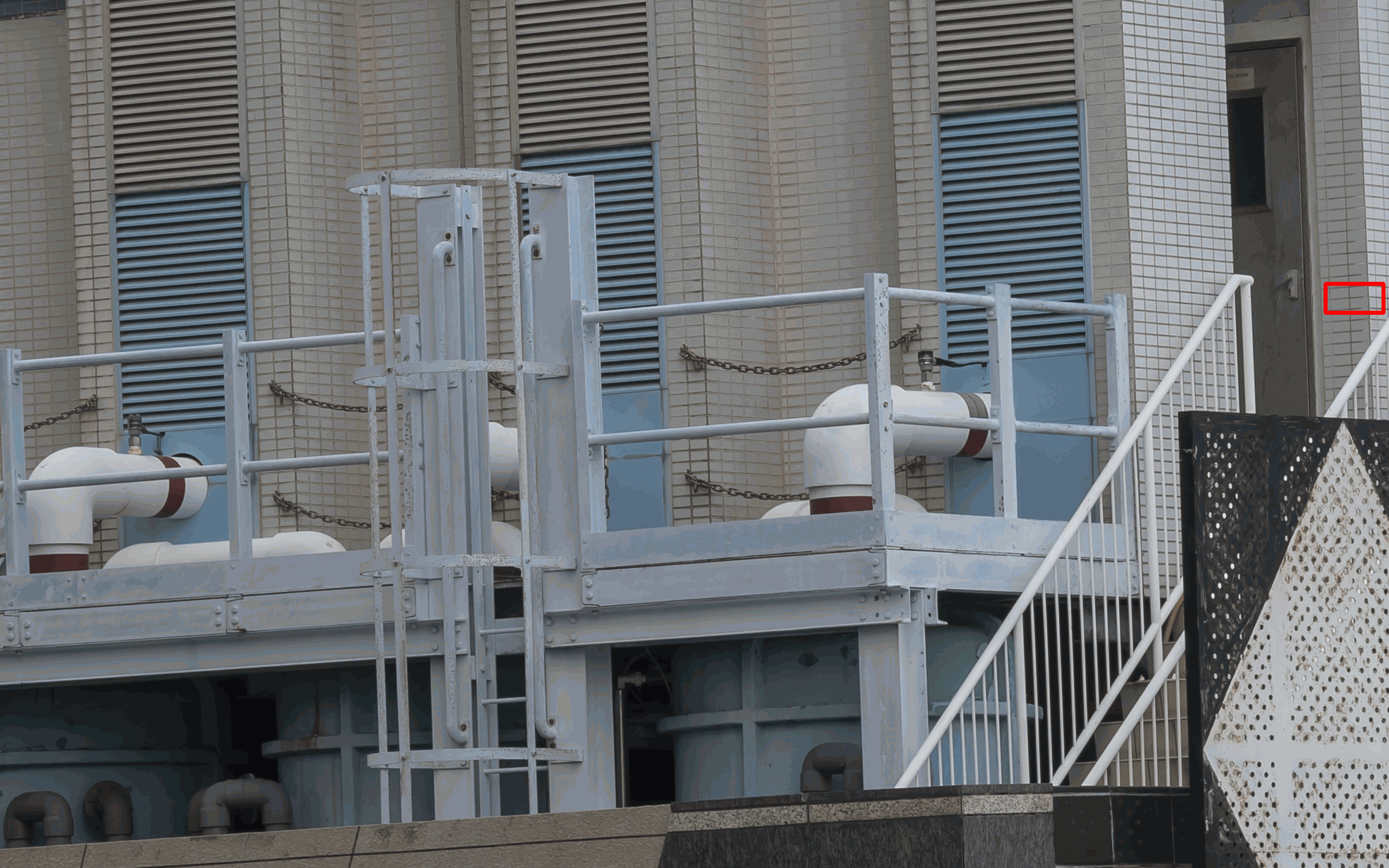} \\
					\fontsize{6.5pt}{1pt}\selectfont RealSRv3 ($\times 4$): Nikon044 \\
				\end{tabular}
			\end{adjustbox}
			\hspace{-3mm}
			\begin{adjustbox}{valign=t}
				\begin{tabular}{ccc}
					\includegraphics[width=0.09\textwidth, height=0.055\textheight]{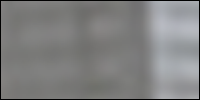} & 
					\hspace{-3mm}
                    \includegraphics[width=0.09\textwidth, height=0.055\textheight]{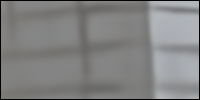} & 
					\hspace{-3mm}
					\includegraphics[width=0.09\textwidth, height=0.055\textheight]{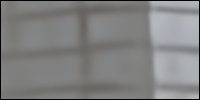} \\
					\fontsize{6.5pt}{1pt}\selectfont LR & \hspace{-3mm}
					\fontsize{6.5pt}{1pt}\selectfont SwinIR-light~\cite{liang2021swinir} & \hspace{-3mm}
                    \fontsize{6.5pt}{1pt}\selectfont ELAN-light~\cite{zhang2022efficient}\\
					\includegraphics[width=0.09\textwidth, height=0.055\textheight]{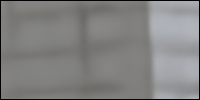} & 
					\hspace{-3mm}
                    \includegraphics[width=0.09\textwidth, height=0.055\textheight]{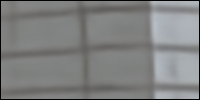} & 
					\hspace{-3mm}
					\includegraphics[width=0.09\textwidth, height=0.055\textheight]{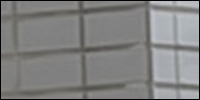} \\
                    \fontsize{6.5pt}{1pt}\selectfont NGswin~\cite{choi2023n} & \hspace{-3mm}
                    \fontsize{6.5pt}{1pt}\selectfont \textbf{Ours} & \hspace{-3mm}
					\fontsize{6.5pt}{1pt}\selectfont Ground Truth \\
				\end{tabular}
			\end{adjustbox}
			\\
			
	\end{tabular} }
	\vspace{-0mm}
	\caption{Qualitative comparisons with existing methods on the real-world test set RealSRv3~\cite{cai2019toward}.}
	\label{realsr-img}
\end{figure*}


\subsubsection{Comparison on Synthesized Datasets} TABLE~\ref{tab:performance} presents quantitative comparisons among various lightweight SR methods at scale factors of \hbox{$ \times 2$}, \hbox{$ \times 3$}, and \hbox{$ \times 4$}. With minimal costs, our DMNet outperforms state-of-the-art methods in terms of PSNR and SSIM on all benchmark test datasets. Specifically, compared to the latest CNN-based method GASSL-B~\cite{wang2023global}, we surpass its PSNR by 0.24dB, 0.26dB, and 0.19dB at three scale factors, respectively, while utilizing only 83$\%$ of its number of parameters and 72$\%$ of its FLOPs. Compared to the latest Transformer-based method NGswin~\cite{choi2023n}, we surpass its PSNR by 0.18dB, 0.14dB, and 0.16dB at three scale factors, respectively, while utilizing only 57$\%$ of its parameters and 82$\%$ of its FLOPs. In Fig.~\ref{visual}, we present a visual comparison of three scale factors. Our method accurately reconstructs high-frequency boundaries of walls, eaves, and letters, highlighting the texture of contours. In contrast, images reconstructed by existing methods appear blurry or incorrect textures. 

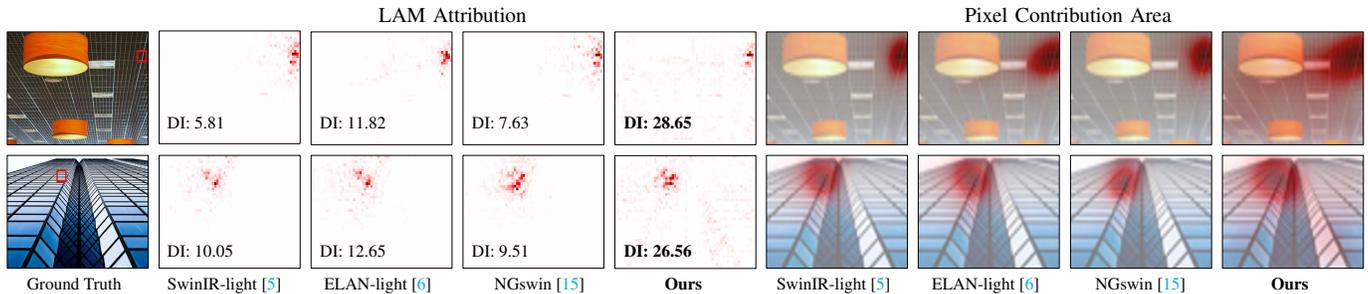
\begin{figure*}[ht]
\vspace{3.5mm}
\begin{overpic}[width=0.999\linewidth]{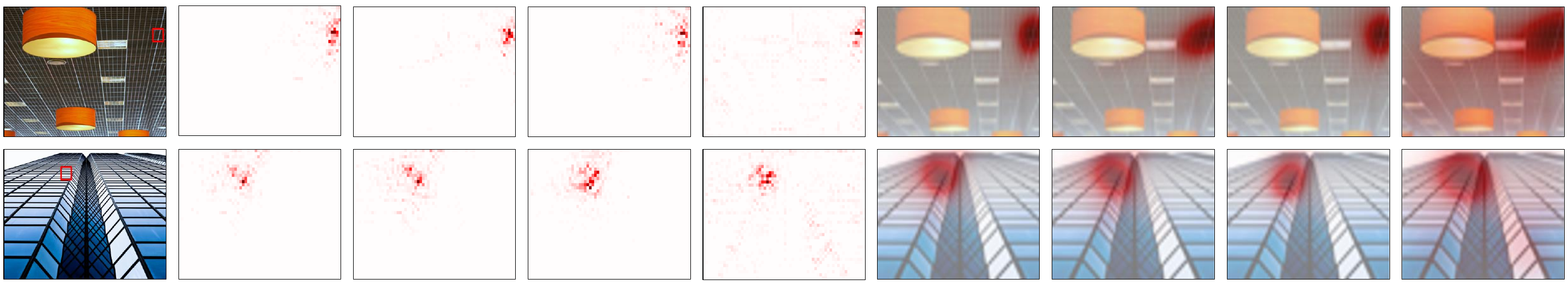}
\put(27.5,18.4){\color{black}{\fontsize{8pt}{1pt}\selectfont LAM Attribution}}
\put(70.5,18.4){\color{black}{\fontsize{8pt}{1pt}\selectfont Pixel Contribution Area}}

\put(1.7,-1.2){\color{black}{\fontsize{6.5pt}{1pt}\selectfont Ground Truth}}
\put(12.0,-1.2){\color{black}{\fontsize{6.5pt}{1pt}\selectfont SwinIR-light~\cite{liang2021swinir}}}
\put(23.5,-1.2){\color{black}{\fontsize{6.5pt}{1pt}\selectfont ELAN-light~\cite{zhang2022efficient}}}
\put(36.0,-1.2){\color{black}{\fontsize{6.5pt}{1pt}\selectfont NGswin~\cite{choi2023n}}}
\put(48.5,-1.2){\color{black}{\fontsize{6.5pt}{1pt}\selectfont \bf{Ours}}}

\put(56.7,-1.2){\color{black}{\fontsize{6.5pt}{1pt}\selectfont SwinIR-light~\cite{liang2021swinir}}}
\put(68.1,-1.2){\color{black}{\fontsize{6.5pt}{1pt}\selectfont ELAN-light~\cite{zhang2022efficient}}}
\put(80.1,-1.2){\color{black}{\fontsize{6.5pt}{1pt}\selectfont NGswin~\cite{choi2023n}}}
\put(93.0,-1.2){\color{black}{\fontsize{6.5pt}{1pt}\selectfont \bf{Ours}}}

\put(12.0,1.2){\color{black}{\fontsize{6.5pt}{1pt}\selectfont DI: 10.05}}
\put(23.2,1.2){\color{black}{\fontsize{6.5pt}{1pt}\selectfont DI: 12.65 }}
\put(34.35,1.2){\color{black}{\fontsize{6.5pt}{1pt}\selectfont DI: 9.51}}
\put(45.5,1.2){\color{black}{\fontsize{6.5pt}{1pt}\selectfont \bf{DI: 26.56}}}

\put(12.0,10.6){\color{black}{\fontsize{6.5pt}{1pt}\selectfont DI: 5.81}}
\put(23.2,10.6){\color{black}{\fontsize{6.5pt}{1pt}\selectfont DI: 11.82 }}
\put(34.35,10.6){\color{black}{\fontsize{6.5pt}{1pt}\selectfont DI: 7.63}}
\put(45.5,10.6){\color{black}{\fontsize{6.5pt}{1pt}\selectfont \bf{DI: 28.65}}}

\end{overpic}
\vspace{1mm}
   \caption{Comparison with existing methods on LAM~\cite{gu2021interpreting} results at the scale of ×4, which indicates the pixel range utilized by networks when inferring results for the portion outlined with the red box. DI quantifies results of LAM~\cite{gu2021interpreting}, with our method achieving the highest DI score and the widest pixel contribution area, indicating its ability to capture broader features.}
\label{fig:interaction}
\vspace{-0mm}
\end{figure*}
\begin{figure*}[h]
	\scriptsize
	\centering
	\scalebox{0.88}{
		\begin{tabular}{lc}
                \begin{adjustbox}{valign=t}
				\begin{tabular}{c}
					\includegraphics[width=0.2\textwidth, height=0.12\textheight]{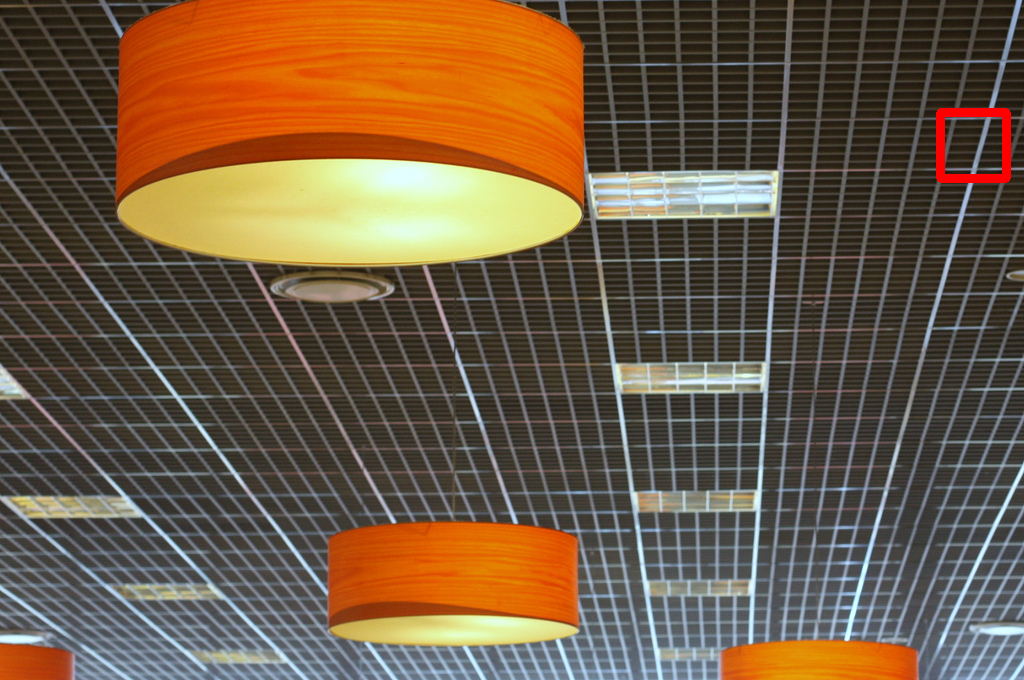} \\
					\fontsize{7.5pt}{1pt}\selectfont Urban100 ($\times 4$): img044 \\
				\end{tabular}
			\end{adjustbox}
			\hspace{-2mm}
			\begin{adjustbox}{valign=t}
				\begin{tabular}{ccc}
					\includegraphics[width=0.09\textwidth, height=0.0525\textheight]{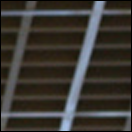} & 
					\hspace{-2mm}
					\includegraphics[width=0.09\textwidth, height=0.0525\textheight]{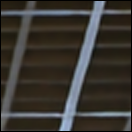} & 
					\hspace{-2mm}
					\includegraphics[width=0.09\textwidth, height=0.0525\textheight]{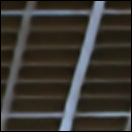} \\
					\fontsize{7.5pt}{1pt}\selectfont LR & \hspace{-2mm}
				  \fontsize{7.5pt}{1pt}\selectfont SwinIR-light & \hspace{-2mm}
					\fontsize{7.5pt}{1pt}\selectfont ELAN-light \\

					\includegraphics[width=0.09\textwidth, height=0.0525\textheight]{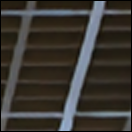} & 
					\hspace{-2mm}
					\includegraphics[width=0.09\textwidth, height=0.0525\textheight]{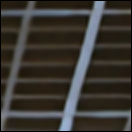} &
					\hspace{-2mm}
					\includegraphics[width=0.09\textwidth, height=0.0525\textheight]{Patch/supp_iter_urbanx4_044/resultcrop_img_044_HR_x4.png} \\
					\fontsize{7.5pt}{1pt}\selectfont NGSwin & \hspace{-2mm}
					\fontsize{7.5pt}{1pt}\selectfont \textbf{Ours} & \hspace{-2mm}
					\fontsize{7.5pt}{1pt}\selectfont Ground Truth \\
				\end{tabular}
			\end{adjustbox}

            \begin{adjustbox}{valign=t}
				\begin{tabular}{c}
					\includegraphics[width=0.2\textwidth, height=0.12\textheight]{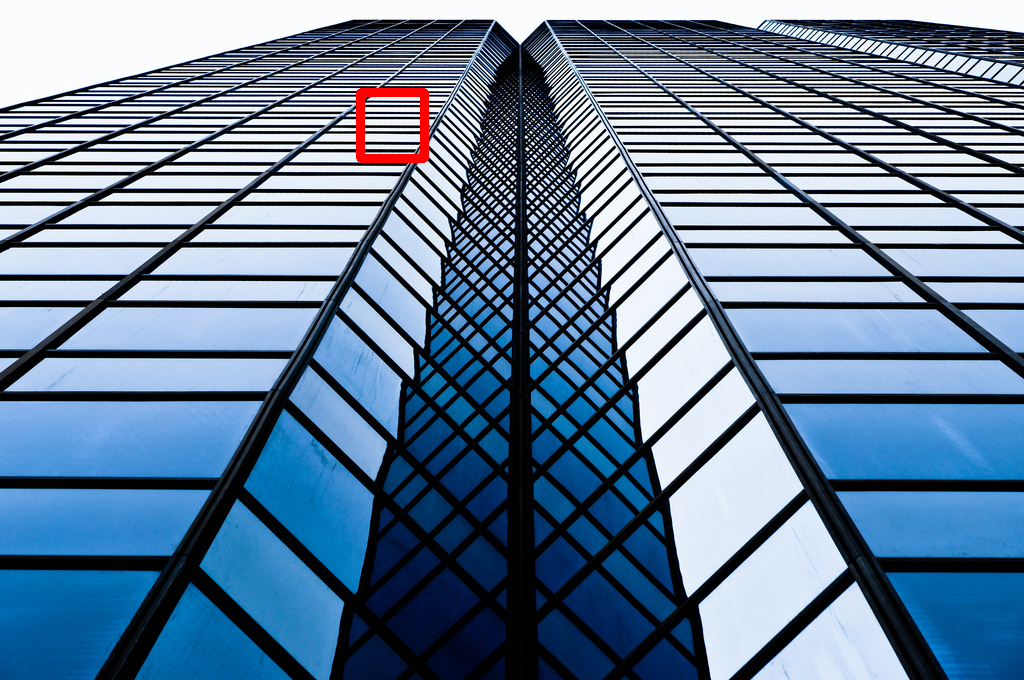} \\
					\fontsize{7.5pt}{1pt}\selectfont Urban100 ($\times 4$): img067 \\
				\end{tabular}
			\end{adjustbox}
			\hspace{-2mm}
			\begin{adjustbox}{valign=t}
				\begin{tabular}{ccc}
					\includegraphics[width=0.09\textwidth, height=0.0525\textheight]{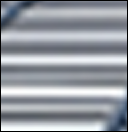} & 
					\hspace{-2mm}
					\includegraphics[width=0.09\textwidth, height=0.0525\textheight]{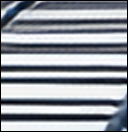} & 
					\hspace{-2mm}
					\includegraphics[width=0.09\textwidth, height=0.0525\textheight]{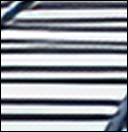} \\
					\fontsize{7.5pt}{1pt}\selectfont LR & \hspace{-2mm}
				  \fontsize{7.5pt}{1pt}\selectfont SwinIR-light & \hspace{-2mm}
					\fontsize{7.5pt}{1pt}\selectfont ELAN-light \\

					\includegraphics[width=0.09\textwidth, height=0.0525\textheight]{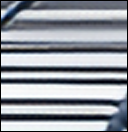} & 
					\hspace{-2mm}
					\includegraphics[width=0.09\textwidth, height=0.0525\textheight]{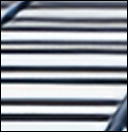} &
					\hspace{-2mm}
					\includegraphics[width=0.09\textwidth, height=0.0525\textheight]{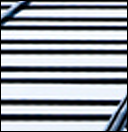} \\
					\fontsize{7.5pt}{1pt}\selectfont NGSwin & \hspace{-2mm}
					\fontsize{7.5pt}{1pt}\selectfont \textbf{Ours} & \hspace{-2mm}
					\fontsize{7.5pt}{1pt}\selectfont Ground Truth \\
				\end{tabular}
			\end{adjustbox}
			
	\end{tabular} }
	\vspace{-0mm}
	\caption{Qualitative comparisons with existing Transformer-based methods at the scale of ×4.}
	\label{supp_inter_visual}
\end{figure*}

Moreover, we contrast our method with existing methods using the interpretability analysis tool LAM~\cite{gu2021interpreting} to examine the pixel range utilized for SR reconstruction. LAM can illustrate the correlation between each pixel and the patch outlined in red during the inference process. The diffusion index (DI) and pixel contribution regions are depicted in Fig~\ref{fig:interaction}, where a higher DI or more pixel contribution regions signify a broader range of pixels engaged in the model. Our method's higher DI value and broader pixel contribution area compared to existing methods indicate that it captures more contextual information, showcasing the effectiveness of utilizing wavelet information and Fourier supervision simultaneously. In addition, Figure~\ref{supp_inter_visual} shows corresponding visual comparisons. Our method not only has higher LAM scores but also has clearer SR reconstructions in corresponding image red patches than existing methods. 

\subsubsection{Comparison on Real-world Datasets} We conduct experiments on RealSRv3~\cite{cai2019toward}, which generates LR and HR image pairs of real scenes by modulating camera focal lengths. TABLE~\ref{tab:realsr} presents the results of quantitative comparisons with IMDN~\cite{hui2019lightweight}, SwinIR-light~\cite{liang2021swinir}, ELAN-light~\cite{zhang2022efficient}, NGswin~\cite{choi2023n}, and MambaIR~\cite{guo2025mambair} at scale factors of \hbox{$ \times 4$}. Our method reconstructs images with higher pixel similarity metrics, such as PSNR and SSIM, as well as perceptual metrics, like LPIPS, at lower computational costs compared to existing lightweight methods. Additionally, including Nikon and Canon datasets, Fig.~\ref{realsr-img} compares visual results in real-world datasets, where our method reconstructs sharper image textures and closely resembles the ground truth compared to existing methods.

\subsubsection{Model Efficiency Tradeoffs} TABLE~\ref{tab:trans_mamba_performance} shows a comprehensive comparison with existing lightweight Transformer- and Mamba-based SR methods, including the number of Parmas, number of FLOPs, inference time, GPU memory, and performance at the scale of $\times 2$. Our method boasts the smallest model size and GPU memory, the fastest inference, and optimal PSNR across various scenarios. Notably, our WMT enables interactions between different frequency subbands within a sensory field downsampled by a factor of 2, significantly accelerating inference by 1.6× to 15.4× compared to existing Transformer- and Mamba-based methods. 


\subsection{Ablation Studies}

\subsubsection{Ablation Study on WMA Configurations} 
\begin{figure}[t]
\begin{overpic}[width=0.99\linewidth]
{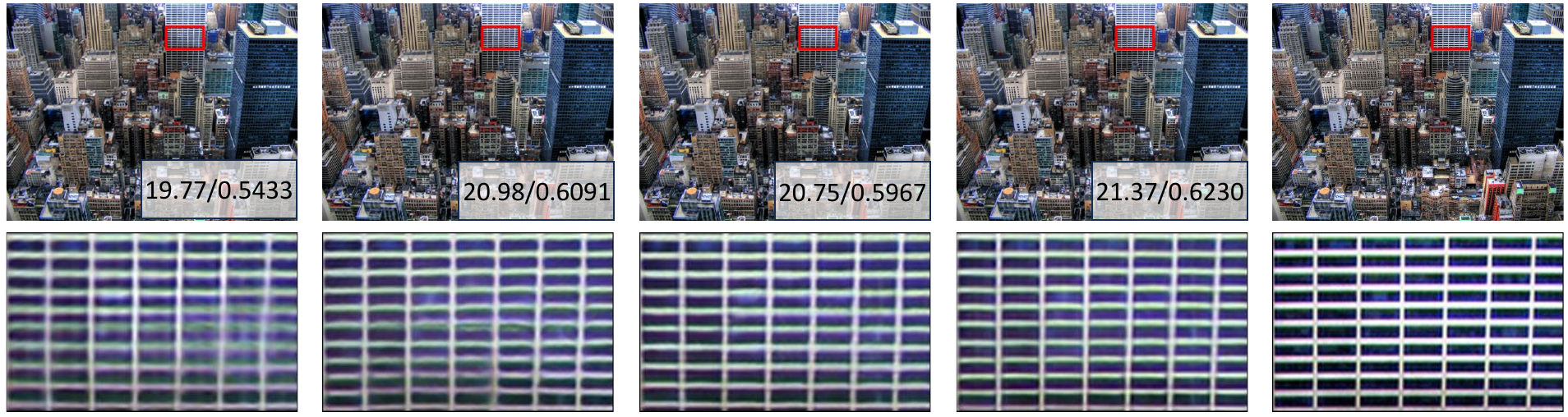}
\put(5.0,-3){\color{black}{\fontsize{7pt}{1pt}\selectfont Baseline}}
\put(23.3,-3){\color{black}{\fontsize{7pt}{1pt}\selectfont w/ Fourier}}
\put(43.2,-3){\color{black}{\fontsize{7pt}{1pt}\selectfont w/ Wavelet}}
\put(67.6,-3){\color{black}{\fontsize{7pt}{1pt}\selectfont \textbf{Ours}}}
\put(82.7,-3){\color{black}{\fontsize{7pt}{1pt}\selectfont Ground truth}}
\end{overpic}
\vspace{4.5mm}
   \caption{Fourier and Wavelet domain information facilitate the reconstruction of accurate overall structures (PSNR and SSIM) and sharp SR textures. }
\label{fig:ablation_comparison}
\vspace{-0mm}
\end{figure}
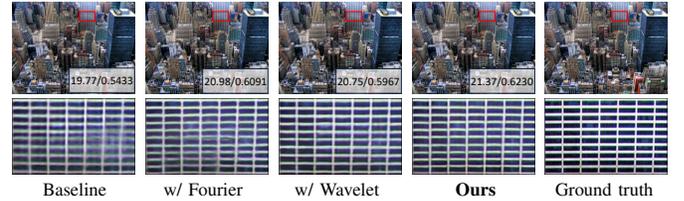

In TABLE~\ref{tab:WMT}, we conduct an
ablation study to investigate the effect of various components in WMA, including the following variants: 

\emph{\textbf{Case1:}} The selection of frequency domain includes the wavelet domain after the wavelet transform and the Fourier domain after the Fourier transform. \emph{\textbf{Case2:}} The selection of frequency loss for supervising the frequency reconstruction includes Wavelet loss and Fourier loss as described in subsection~\ref{Loss}. \emph{\textbf{Case3:}} The selection of internal module structures of WMA includes a dynamic branch responsible for enhancing frequency details. By combining Case1 and Case2, it is evident that selecting the wavelet transform as the frequency decomposition function, with Fourier loss supervision, is the optimal choice for achieving accurate SR results at lower costs. As shown in Fig~\ref{fig:ablation_comparison}, wavelet and Fourier information facilitate the recovery of high-frequency features and overall structure, respectively. Moreover, the dynamic weight within WMA aids in SR, resulting in a PSNR gain of 0.06dB with an increase in the number of Parmas of less than 4k. 


\subsubsection{Ablation Study on DMNet Configurations} 
\begin{figure*}
\begin{overpic}[width=0.97\linewidth]{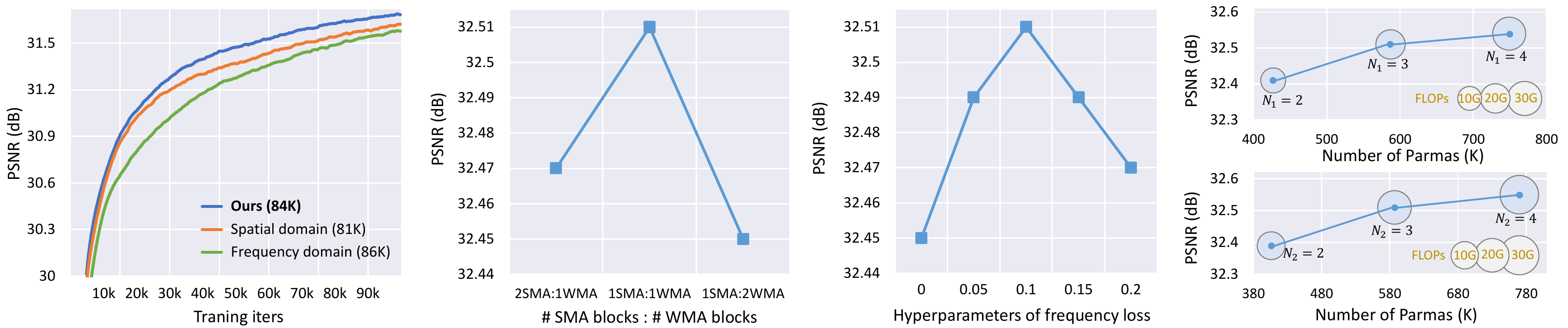}
\put(5.5,-2.5){\color{black}{\fontsize{7pt}{1pt}\selectfont (a) PSNR lines during training.}}
\put(28.8,-2.5){\color{black}{\fontsize{6.8pt}{1pt}\selectfont (b) Ratio of number of WMA and SMA.}}
\put(58.1,-2.5){\color{black}{\fontsize{7pt}{1pt}\selectfont (c) Hyperparameter \hbox{$\lambda $}.}}
\put(81.6,-2.5){\color{black}{\fontsize{7pt}{1pt}\selectfont (d) Values of \hbox{${N_1}$} and \hbox{${N_2}$}.}}
\end{overpic}
\vspace{5.5mm}
   \caption{Ablation studies of our DMNet at scale factors of \hbox{$ \times 4$} on Set5~\cite{bevilacqua2012low} test set, where all methods in (a) use the loss of Eq.(\ref{eq: 16}). \hbox{$N_1$} and \hbox{$N_2$} refer to the number of SWBlock and SWGroup in Fig~\ref{fig:main}.} 
\label{fig:abalation_DMNet}
\vspace{-0mm}
\end{figure*}
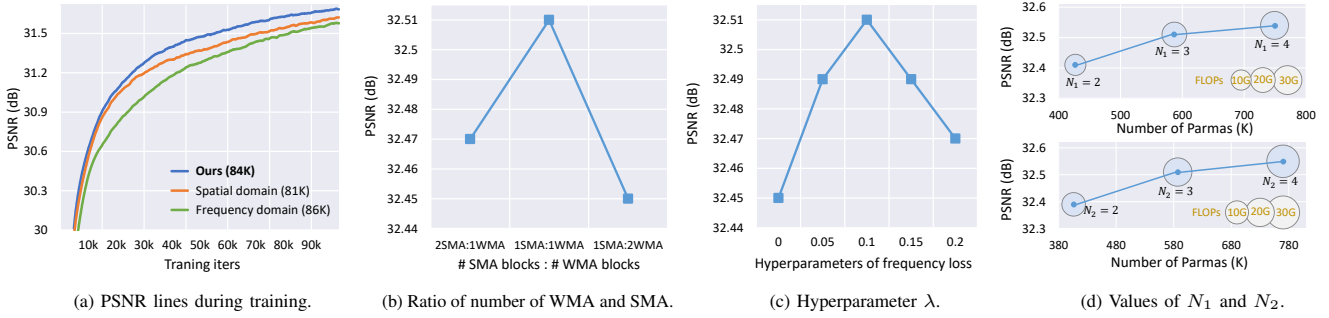

To validate the robustness of our DMNet configuration, we present a series of ablations in Fig~\ref{fig:abalation_DMNet}, which includes the following variants: 

\emph{\textbf{Case1:}} The benefits of focusing on both spatial and frequency domain information. As depicted in Fig~\ref{fig:abalation_DMNet} (a), when supervised by both spatial and frequency losses, concentrating on both spatial and frequency domain feature reconstruction leads to faster convergence and superior performance compared to focusing solely on the spatial or frequency domain without incurring too much computational consumption.

\begin{table*}[t!]
\tiny
\setlength\tabcolsep{4pt}
\centering
\vspace{0mm}
    \caption{Ablation on configurations of wavelet-domain modulation self-attention (WMA) at scale factors of \hbox{$ \times 4$}. ``Wavelet domain'' and ``Fourier domain'' refer to the choice of frequency domain chose within WMA. ``Wavelet loss'' and ``Fourier loss'' refer to the choice of frequency loss. ``Dynamic'' refers to the dynamic weight in WMA.}
\vspace{-0mm}
\label{tab:WMT}
\resizebox{0.96\textwidth}{!}{
\begin{tabular}{l||cc|cc|c|l|l|cc}
\toprule
\rowcolor{lightgray}
& \multicolumn{2}{c|}{Frequency Domain} 
& \multicolumn{2}{c|}{Frequency Loss} 
& \multicolumn{1}{c|}{Module Structure} 
& & & \multicolumn{2}{c}{Set5} 
\\ 
\cmidrule{2-6}
\cmidrule{9-10}
    \rowcolor{lightgray}
    \multicolumn{1}{l||}{\multirow{-2}{*}{Methods}} 
    & Wavelet domain  & Fourier domain    
    & Wavelet loss  & Fourier loss     
    & Dynamic
    & \multicolumn{1}{c|}{\multirow{-2}{*}{Params}}
    & \multicolumn{1}{c|}{\multirow{-2}{*}{FLOPs}}
    & PSNR  & SSIM  
    \\ 
    \hline
    \emph{w/o} Dynamic  & \Checkmark   &   &      & \Checkmark  & \XSolidBrush    & 79.8K  & 4.15G  & 31.57 & 0.8853  \\
    Model1  &    & \Checkmark  & \Checkmark     &   & \Checkmark    & 84.8K  & 4.82G  & 31.57  & 0.8846  \\
    Model2  &    & \Checkmark  &      & \Checkmark  & \Checkmark    & 84.8K  & 4.82G  & 31.56 & 0.8842 \\
    Model3  & \Checkmark   &   & \Checkmark     &   & \Checkmark    & 83.6K  & 4.19G  & 31.54  & 0.8837  \\
    \bf{Ours}   & \Checkmark   &   &      & \Checkmark  & \Checkmark    & 83.6K  & 4.19G  & \bf{31.63} & \bf{0.8858} \\

    \bottomrule
\end{tabular}}
\end{table*}

\emph{\textbf{Case2:}} The optimal ratio of modules that concentrate on the spatial domain to those that focus on the frequency domain. In our DMNet, employing SMA for the spatial domain and WMA for the frequency domain, we aim to determine the optimal ratio between the two modules. 
Fig~\ref{fig:abalation_DMNet} (b) demonstrates that the one-to-one ratio currently employed in our method is optimal. It is worth noting that the stacking order of spatial and frequency blocks has limited impact on overall performance. However, parallel architectures demand carefully designed frequency–spatial fusion mechanisms. For effectiveness and design simplicity, we adopt a sequential structure where spatial processing precedes frequency modeling.

\emph{\textbf{Case3:}} Ablation study on the hyperparameter \hbox{$\lambda $} of the frequency loss. Fig~\ref{fig:abalation_DMNet} (c) shows that incorporating frequency loss supervision enhances the model's performance, with the optimal performance achieved at \hbox{$\lambda  = 0.1$}. Deviating too much or too little from this value can negatively impact performance, thus justifying our choice of \hbox{$\lambda  = 0.1$} for our DMNet. 

\begin{figure}[htbp]
    \centering
    \begin{minipage}[t]{0.24\textwidth}
        \centering
        \begin{overpic}[width=\linewidth]{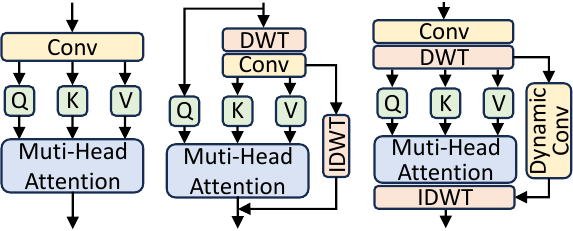}
            \put(-1,-5){\color{black}{\fontsize{6pt}{1pt}\selectfont (a) Restormer}}
            \put(32.0,-5){\color{black}{\fontsize{6pt}{1pt}\selectfont (b) Wave-Vit}}
            \put(72.0,-5){\color{black}{\fontsize{6pt}{1pt}\selectfont (c) \textbf{Ours}}}
        \end{overpic}
    \end{minipage}%
    \hfill%
    \begin{minipage}[t]{0.22\textwidth}
        \vspace{-18.5mm}
        \centering
        \setlength\tabcolsep{3pt}
        \renewcommand{\arraystretch}{1.1}
        \resizebox{\linewidth}{!}{
        \begin{tabular}{l||ccc}
            \toprule
            \rowcolor{lightgray}
            Modules & (a) & (b) & \textbf{Ours} \\
            \hline
            Parmas$\downarrow$ & \textbf{80.9K} & 102.7K & 83.6K \\
            FLOPs$\downarrow$  & 4.7G & 7.1G & \textbf{4.2G} \\
            Memory$\downarrow$ & 2.12G & 9.66G & \textbf{1.96G} \\
            PSNR$\uparrow$     & 31.56 & 31.39 & \textbf{31.63} \\
            SSIM$\uparrow$     & 0.8844 & 0.8811 & \textbf{0.8858} \\
            \bottomrule
        \end{tabular}}
    \end{minipage}
    \caption{Comparison of our proposed WMA and related self-attention modules from Restormer~\cite{zamir2022restormer} and Wave-Vit~\cite{yao2022wave}.}
    \label{fig:wma_comparison}
\end{figure}

\emph{\textbf{Case4:}} 
Impact of values of \hbox{${N_1}$} and \hbox{${N_2}$} on performance. In our DMNet in Fig~\ref{fig:main}, we partition the combined SMT and WMT structure into distinct SWBlocks and SWGroups, which are then interconnected using residual connections. Fig~\ref{fig:abalation_DMNet} (d) investigates the optimal selection of the number of \hbox{${N_1}$} and \hbox{${N_2}$}. As \hbox{$N_1$} and \hbox{$N_2$} increase from 2 to 3, there is a rise in PSNR of nearly 0.1 dB, accompanied by a certain increase in computational consumption. However, when \hbox{${N_1}$} and \hbox{${N_2}$} are increased from 3 to 4, the gain in PSNR is relatively smaller, even though the computational consumption increases further. Consequently, in our DMNet, we choose \hbox{${N_1} = {N_2} = 3$}.

\subsubsection{Comparison of WMA and Existing Self-attentions} As shown in Fig~\ref{fig:wma_comparison}, We give structural differences between ours and relevant self-attentions. Compared to Restormer~\cite{zamir2022restormer} and Wave-Vit~\cite{yao2022wave}, which have a computational complexity of \hbox{$\mathcal{O}\left( {C^2}N \right)$} and \hbox{$\mathcal{O}\left( {N^2}C/16 \right)$}, respectively, our computational complexity is smaller, only \hbox{$\mathcal{O}\left( {C^2}N/4 \right)$}, where \hbox{$N = H \times W$}, \hbox{$H$}, \hbox{$W$}, \hbox{$C$} are the height, width, and number of channels of the feature, respectively. Additionally, our WMA can achieve SR more efficiently than these methods, including computational costs, GPU memory in training, and performance on the Set5 test set. Hence, our WMA is a better solution for frequency-based SR than existing self-attentions.

\begin{figure*}[ht]
\begin{overpic}[width=0.99\linewidth]{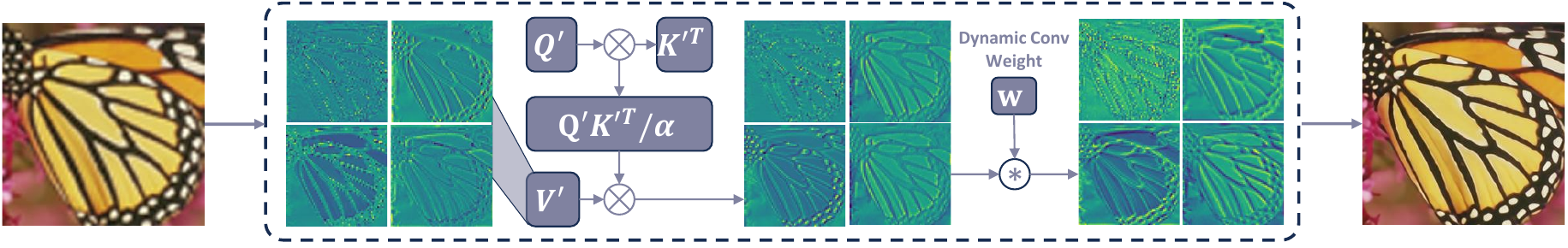}
\end{overpic}
\vspace{-0mm}
   \caption{Visualization of features within our WMA. First, self-attention learns relationships between high and low-frequency information in the wavelet domain, enabling mutual enhancement to reconstruct global dependencies. Then, dynamic convolution adaptively captures the details of each frequency component to refine local features.}
\label{fig:WMA_fea_visual}
\vspace{-0mm}
\end{figure*}

\subsubsection{Effectiveness of WMT} We compare our WMT with existing frequency-based SR modules, including methods from CNN-based and transformer-based methods for processing wavelet and Fourier features. Notably, these frequency-domain processing modules originate from general SR methods, as lightweight SR currently lacks dedicated modules for frequency exploration. As shown in TABLE~\ref{tab:ablation_cnn_Transformer}, compared to these modules, our WMT is more efficient, achieving higher PSNR/SSIM with lower computational cost and faster inference. 

Next, we present a feature map visualization of the self-attention part (WMA) within the WMT interior to highlight the role of wavelet transform combined with self-attention and dynamic convolution. As shown in Fig~\ref{fig:WMA_fea_visual}, for input features with one low-frequency component and three high-frequency components, self-attention improves global textures by capturing long-range dependencies, leading to a more coherent overall structure. However, it fails to fully capture subtle low-frequency details. Dynamic convolution complements this by adaptively refining these details, ensuring that both global structures and fine-grained local features are accurately restored, thus enhancing the overall quality of feature maps.

\section{Limitations and Future Works}~\label{Limitations}
While our proposed DMNet achieves competitive results in lightweight SR, there are several areas for improvement. The model could be further optimized for extremely LR images, where the current SR performance may not be as robust. Additionally, more advanced fusion techniques between frequency and spatial domains could enhance the overall feature extraction and reconstruction process. The scalability of the model for larger datasets and real-time applications also remains a challenge. In the future, we plan to explore techniques for generating real LR data~\cite{peng2024towards}, efficient state space models such as Mamba~\cite{guo2025mambair}, advanced architectures like Gaussian Splatting~\cite{peng2025pixel}, and adaptive frequency-spatial fusion strategies~\cite{xiao2024frequency} to further enhance model robustness and generalization in real-world SR scenarios, including real degradations and arbitrary scale task.

\begin{table}[t!]
\tiny
\setlength\tabcolsep{4.5pt}
\centering
\vspace{0mm}
    \caption{Ablation study on existing frequency-based modules, including WTM in JWSGN~\cite{zou2022joint}, FSAS in FFTformer~\cite{kong2023efficient}, WaveMix in WaveMixSR~\cite{jeevan2024wavemixsr}, and WAT in SODA-SR~\cite{ai2024uncertainty}, with our proposed WMT, on the $\times 4$ Set5~\cite{matsui2017sketch} test set.}
\vspace{-0mm}
\label{tab:ablation_cnn_Transformer}
\resizebox{0.48\textwidth}{!}{
\begin{tabular}{l||l|l|l||l}
\toprule
\rowcolor{lightgray}
    \rowcolor{lightgray}
    \multicolumn{1}{l||}{\multirow{-1}{*}{Modules}} 
    & \multicolumn{1}{c|}{\multirow{-1}{*}{Params}}
    & \multicolumn{1}{c|}{\multirow{-1}{*}{FLOPs}}
    & \multicolumn{1}{c||}{\multirow{-1}{*}{Speed}}
    & \multicolumn{1}{c}{\multirow{-1}{*}{PSNR / SSIM}}

    \\ 
    \hline
    WTM~\cite{zou2022joint}        & 134.3K & 5.49G    & 90ms & 31.43 / 0.8840 \\ 
    FSAS~\cite{kong2023efficient}      & 137.2K & 7.90G     & 35ms & 31.51 / 0.8845 \\
    WaveMix~\cite{jeevan2024wavemixsr}      & 146.2K & 5.76G    & 46ms & 31.41 / 0.8839 \\
    WAT~\cite{ai2024uncertainty}      & 126.3K & 5.51G    & 129ms & 31.55 / 0.8849 \\
    \textbf{Ours}      & \textbf{83.6K} & \textbf{4.19G}    & 26ms  & \textbf{31.63} / \textbf{0.8858} \\ 

    \bottomrule
\end{tabular}}
\end{table}
\hspace{-0mm}

\section{Conclusion}~\label{Conclusion}
This paper presents a dual-domain modulation network for lightweight SR. We introduce SMT and WMT to address reconstruction in the spatial and frequency domains to enhance performance while maintaining lightweight. Meanwhile, the above process is supervised through image reconstruction and Fourier loss. We show that by combining wavelet features with Fourier supervision, our WMT can reconstruct high-frequency features while maintaining overall quality. Further, WMT as a frequency domain exploration module designed for lightweight SR, effectively enhances SR with minimal overhead. Through extensive experiments on both synthesized and real datasets, we show our method achieves superior performance with lower complexity, surpassing state-of-the-art lightweight SR methods. 

As this work is mainly based on synthetic datasets and fixed SR scales, we plan to explore advanced architectures such as Mamba and 3D Gaussian Splatting, and design adaptive frequency-spatial fusion and degradation-aware strategies to improve robustness in real-world SR scenarios.

\bibliographystyle{IEEEtran}
\bibliography{sample-base}

\begin{thebibliography}{10}
\providecommand{\url}[1]{#1}
\csname url@samestyle\endcsname
\providecommand{\newblock}{\relax}
\providecommand{\bibinfo}[2]{#2}
\providecommand{\BIBentrySTDinterwordspacing}{\spaceskip=0pt\relax}
\providecommand{\BIBentryALTinterwordstretchfactor}{4}
\providecommand{\BIBentryALTinterwordspacing}{\spaceskip=\fontdimen2\font plus
\BIBentryALTinterwordstretchfactor\fontdimen3\font minus \fontdimen4\font\relax}
\providecommand{\BIBforeignlanguage}[2]{{%
\expandafter\ifx\csname l@#1\endcsname\relax
\typeout{** WARNING: IEEEtran.bst: No hyphenation pattern has been}%
\typeout{** loaded for the language `#1'. Using the pattern for}%
\typeout{** the default language instead.}%
\else
\language=\csname l@#1\endcsname
\fi
#2}}
\providecommand{\BIBdecl}{\relax}
\BIBdecl

\bibitem{zhou2023srformer}
Y.~Zhou, Z.~Li, C.-L. Guo, S.~Bai, M.-M. Cheng, and Q.~Hou, ``Srformer: Permuted self-attention for single image super-resolution,'' in \emph{ICCV}, 2023, pp. 12\,780--12\,791.

\bibitem{guo2025mambair}
H.~Guo, J.~Li, T.~Dai, Z.~Ouyang, X.~Ren, and S.-T. Xia, ``Mambair: A simple baseline for image restoration with state-space model,'' in \emph{ECCV}.\hskip 1em plus 0.5em minus 0.4em\relax Springer, 2024, pp. 222--241.

\bibitem{georgescu2023multimodal}
M.-I. Georgescu, R.~T. Ionescu, A.-I. Miron, O.~Savencu, N.-C. Ristea, N.~Verga, and F.~S. Khan, ``Multimodal multi-head convolutional attention with various kernel sizes for medical image super-resolution,'' in \emph{WACV}, 2023, pp. 2195--2205.

\bibitem{xiao2023ediffsr}
Y.~Xiao, Q.~Yuan, K.~Jiang, J.~He, X.~Jin, and L.~Zhang, ``Ediffsr: An efficient diffusion probabilistic model for remote sensing image super-resolution,'' \emph{IEEE Trans. Geosci. Remote Sens.}, vol.~62, pp. 1--14, 2023.

\bibitem{jiang2020dual}
K.~Jiang, Z.~Wang, P.~Yi, T.~Lu, J.~Jiang, and Z.~Xiong, ``Dual-path deep fusion network for face image hallucination,'' \emph{IEEE Trans. Neural Netw. Learn. Syst.}, vol.~33, no.~1, pp. 378--391, 2020.

\bibitem{noguchi2024scene}
C.~Noguchi, S.~Fukuda, and M.~Yamanaka, ``Scene text image super-resolution based on text-conditional diffusion models,'' in \emph{WACV}, 2024, pp. 1485--1495.

\bibitem{lim2017enhanced}
B.~Lim, S.~Son, H.~Kim, S.~Nah, and K.~Mu~Lee, ``Enhanced deep residual networks for single image super-resolution,'' in \emph{CVPRW}, 2017, pp. 136--144.

\bibitem{dai2019second}
T.~Dai, J.~Cai, Y.~Zhang, S.-T. Xia, and L.~Zhang, ``Second-order attention network for single image super-resolution,'' in \emph{CVPR}, 2019, pp. 11\,065--11\,074.

\bibitem{liang2021swinir}
J.~Liang, J.~Cao, G.~Sun, K.~Zhang, L.~Van~Gool, and R.~Timofte, ``Swinir: Image restoration using swin transformer,'' in \emph{ICCVW}, 2021, pp. 1833--1844.

\bibitem{zhang2022efficient}
X.~Zhang, H.~Zeng, S.~Guo, and L.~Zhang, ``Efficient long-range attention network for image super-resolution,'' in \emph{ECCV}.\hskip 1em plus 0.5em minus 0.4em\relax Springer, 2022, pp. 649--667.

\bibitem{wang2024camixersr}
Y.~Wang, Y.~Liu, S.~Zhao, J.~Li, and L.~Zhang, ``Camixersr: Only details need more" attention",'' in \emph{CVPR}, 2024, pp. 25\,837--25\,846.

\bibitem{gu2021interpreting}
J.~Gu and C.~Dong, ``Interpreting super-resolution networks with local attribution maps,'' in \emph{CVPR}, 2021, pp. 9199--9208.

\bibitem{matsui2017sketch}
Y.~Matsui, K.~Ito, Y.~Aramaki, A.~Fujimoto, T.~Ogawa, T.~Yamasaki, and K.~Aizawa, ``Sketch-based manga retrieval using manga109 dataset,'' \emph{Multimed. Tools Appl.}, vol.~76, pp. 21\,811--21\,838, 2017.

\bibitem{bevilacqua2012low}
M.~Bevilacqua, A.~Roumy, C.~Guillemot, and M.~L. Alberi-Morel, ``Low-complexity single-image super-resolution based on nonnegative neighbor embedding,'' in \emph{BMVC}, vol. 135, 2012, pp. 1--10.

\bibitem{zeyde2012single}
R.~Zeyde, M.~Elad, and M.~Protter, ``On single image scale-up using sparse-representations,'' in \emph{ICCS}.\hskip 1em plus 0.5em minus 0.4em\relax Springer, 2012, pp. 711--730.

\bibitem{gao2022feature}
G.~Gao, W.~Li, J.~Li, F.~Wu, H.~Lu, and Y.~Yu, ``Feature distillation interaction weighting network for lightweight image super-resolution,'' in \emph{AAAI}, vol.~36, no.~1, 2022, pp. 661--669.

\bibitem{hui2019lightweight}
Z.~Hui, X.~Gao, Y.~Yang, and X.~Wang, ``Lightweight image super-resolution with information multi-distillation network,'' in \emph{ACMMM}, 2019, pp. 2024--2032.

\bibitem{sun2023hybrid}
B.~Sun, Y.~Zhang, S.~Jiang, and Y.~Fu, ``Hybrid pixel-unshuffled network for lightweight image super-resolution,'' in \emph{AAAI}, vol.~37, no.~2, 2023, pp. 2375--2383.

\bibitem{choi2023n}
H.~Choi, J.~Lee, and J.~Yang, ``N-gram in swin transformers for efficient lightweight image super-resolution,'' in \emph{CVPR}, 2023, pp. 2071--2081.

\bibitem{li2022efficient}
W.~Li, J.~Li, G.~Gao, W.~Deng, J.~Yang, G.-J. Qi, and C.-W. Lin, ``Efficient image super-resolution with feature interaction weighted hybrid network,'' \emph{IEEE Trans. Multimedia}, 2024.

\bibitem{li2023dlgsanet}
X.~Li, J.~Dong, J.~Tang, and J.~Pan, ``Dlgsanet: lightweight dynamic local and global self-attention networks for image super-resolution,'' in \emph{ICCV}, 2023, pp. 12\,792--12\,801.

\bibitem{zhang2025hit}
X.~Zhang, Y.~Zhang, and F.~Yu, ``Hit-sr: Hierarchical transformer for efficient image super-resolution,'' in \emph{ECCV}.\hskip 1em plus 0.5em minus 0.4em\relax Springer, 2024, pp. 483--500.

\bibitem{wang2023global}
H.~Wang, Y.~Zhang, C.~Qin, L.~Van~Gool, and Y.~Fu, ``Global aligned structured sparsity learning for efficient image super-resolution,'' \emph{IEEE Trans. Pattern Anal. Mach. Intell.}, vol.~45, no.~9, pp. 10\,974--10\,989, 2023.

\bibitem{sun2022shufflemixer}
L.~Sun, J.~Pan, and J.~Tang, ``Shufflemixer: An efficient convnet for image super-resolution,'' in \emph{NeurIPs}, vol.~35, 2022, pp. 17\,314--17\,326.

\bibitem{kong2023efficient}
L.~Kong, J.~Dong, J.~Ge, M.~Li, and J.~Pan, ``Efficient frequency domain-based transformers for high-quality image deblurring,'' in \emph{CVPR}, 2023, pp. 5886--5895.

\bibitem{zou2022joint}
W.~Zou, L.~Chen, Y.~Wu, Y.~Zhang, Y.~Xu, and J.~Shao, ``Joint wavelet sub-bands guided network for single image super-resolution,'' \emph{IEEE Trans. Multimedia}, 2022.

\bibitem{sun2023spatially}
L.~Sun, J.~Dong, J.~Tang, and J.~Pan, ``Spatially-adaptive feature modulation for efficient image super-resolution,'' in \emph{ICCV}, 2023, pp. 13\,190--13\,199.

\bibitem{li2023fsr}
J.~Li, T.~Dai, M.~Zhu, B.~Chen, Z.~Wang, and S.-T. Xia, ``Fsr: A general frequency-oriented framework to accelerate image super-resolution networks,'' in \emph{AAAI}, vol.~37, no.~1, 2023, pp. 1343--1350.

\bibitem{jiang2023fabnet}
X.~Jiang, N.~Wang, J.~Xin, K.~Li, X.~Yang, J.~Li, X.~Wang, and X.~Gao, ``Fabnet: Frequency-aware binarized network for single image super-resolution,'' \emph{IEEE Trans. Image Process.}, vol.~32, pp. 6234--6247, 2023.

\bibitem{zheng2025smfanet}
M.~Zheng, L.~Sun, J.~Dong, and J.~Pan, ``Smfanet: A lightweight self-modulation feature aggregation network for efficient image super-resolution,'' in \emph{ECCV}.\hskip 1em plus 0.5em minus 0.4em\relax Springer, 2024, pp. 359--375.

\bibitem{xu2024fdsr}
P.~Xu, Q.~Liu, H.~Bao, R.~Zhang, L.~Gu, and G.~Wang, ``Fdsr: An interpretable frequency division stepwise process based single-image super-resolution network,'' \emph{IEEE Trans. Image Process.}, 2024.

\bibitem{guo2023spatial}
S.~Guo, H.~Yong, X.~Zhang, J.~Ma, and L.~Zhang, ``Spatial-frequency attention for image denoising,'' \emph{arXiv preprint arXiv:2302.13598}, 2023.

\bibitem{li2024efficient}
W.~Li, H.~Guo, X.~Liu, K.~Liang, J.~Hu, Z.~Ma, and J.~Guo, ``Efficient face super-resolution via wavelet-based feature enhancement network,'' in \emph{ACMMM}, 2024, pp. 4515--4523.

\bibitem{ai2024uncertainty}
Y.~Ai, X.~Zhou, H.~Huang, L.~Zhang, and R.~He, ``Uncertainty-aware source-free adaptive image super-resolution with wavelet augmentation transformer,'' in \emph{CVPR}, 2024, pp. 8142--8152.

\bibitem{li2023survey}
W.~Li, M.~Wang, K.~Zhang, J.~Li, X.~Li, Y.~Zhang, G.~Gao, W.~Deng, and C.-W. Lin, ``Survey on deep face restoration: From non-blind to blind and beyond,'' \emph{arXiv preprint arXiv:2309.15490}, 2023.

\bibitem{liu2025causal}
L.~Liu, S.~Sun, S.~Zhi, F.~Shi, Z.~Liu, J.~Heikkil{\"a}, and Y.~Liu, ``A causal adjustment module for debiasing scene graph generation,'' \emph{IEEE Trans. Pattern Anal. Mach. Intell.}, 2025.

\bibitem{li2025saratr}
W.~Li, W.~Yang, Y.~Hou, L.~Liu, Y.~Liu, and X.~Li, ``Saratr-x: Towards building a foundation model for sar target recognition,'' \emph{IEEE Trans. Image Process.}, 2025.

\bibitem{li2024systematic}
J.~Li, Z.~Pei, W.~Li, G.~Gao, L.~Wang, Y.~Wang, and T.~Zeng, ``A systematic survey of deep learning-based single-image super-resolution,'' \emph{ACM Comput. Surv.}, vol.~56, no.~10, pp. 1--40, 2024.

\bibitem{jiang2020hierarchical}
K.~Jiang, Z.~Wang, P.~Yi, and J.~Jiang, ``Hierarchical dense recursive network for image super-resolution,'' \emph{Pattern Recognit.}, vol. 107, p. 107475, 2020.

\bibitem{wang2023ddistill}
Y.~Wang, T.~Su, Y.~Li, J.~Cao, G.~Wang, and X.~Liu, ``Ddistill-sr: Reparameterized dynamic distillation network for lightweight image super-resolution,'' \emph{IEEE Trans. Multimedia}, vol.~25, pp. 7222--7234, 2023.

\bibitem{park2021dynamic}
K.~Park, J.~W. Soh, and N.~I. Cho, ``A dynamic residual self-attention network for lightweight single image super-resolution,'' \emph{IEEE Trans. Multimedia}, vol.~25, pp. 907--918, 2021.

\bibitem{li2023cross}
W.~Li, J.~Li, G.~Gao, W.~Deng, J.~Zhou, J.~Yang, and G.-J. Qi, ``Cross-receptive focused inference network for lightweight image super-resolution,'' \emph{IEEE Trans. Multimedia}, vol.~26, pp. 864--877, 2024.

\bibitem{xiao2024frequency}
Y.~Xiao, Q.~Yuan, K.~Jiang, Y.~Chen, Q.~Zhang, and C.-W. Lin, ``Frequency-assisted mamba for remote sensing image super-resolution,'' \emph{IEEE Trans. Multimedia}, 2024.

\bibitem{xin2020wavelet}
J.~Xin, J.~Li, X.~Jiang, N.~Wang, H.~Huang, and X.~Gao, ``Wavelet-based dual recursive network for image super-resolution,'' \emph{IEEE Trans. Neural Netw. Learn. Syst.}, vol.~33, no.~2, pp. 707--720, 2020.

\bibitem{magid2021dynamic}
S.~A. Magid, Y.~Zhang, D.~Wei, W.-D. Jang, Z.~Lin, Y.~Fu, and H.~Pfister, ``Dynamic high-pass filtering and multi-spectral attention for image super-resolution,'' in \emph{ICCV}, 2021, pp. 4288--4297.

\bibitem{liu2023spectral}
T.~Liu, J.~Cheng, and S.~Tan, ``Spectral bayesian uncertainty for image super-resolution,'' in \emph{CVPR}, 2023, pp. 18\,166--18\,175.

\bibitem{song2024efficient}
J.~Song, A.~Sowmya, and C.~Sun, ``Efficient hybrid feature interaction network for stereo image super-resolution,'' \emph{IEEE Trans. Multimedia}, 2024.

\bibitem{ju2025towards}
Y.~Ju, J.~Xiao, C.~Zhang, H.~Xie, A.~Luo, H.~Zhou, J.~Dong, and A.~C. Kot, ``Towards marine snow removal with fusing fourier information,'' \emph{Inf. Fusion}, vol. 117, p. 102810, 2025.

\bibitem{xiao2025multi}
Y.~Xiao, Q.~Yuan, K.~Jiang, Y.~Chen, S.~Wang, and C.-W. Lin, ``Multi-axis feature diversity enhancement for remote sensing video super-resolution,'' \emph{IEEE Trans. Image Process.}, 2025.

\bibitem{jiang2024fmrnet}
K.~Jiang, J.~Jiang, X.~Liu, X.~Xu, and X.~Ma, ``Fmrnet: Image deraining via frequency mutual revision,'' in \emph{AAAI}, vol.~38, no.~11, 2024, pp. 12\,892--12\,900.

\bibitem{peng2024lightweight}
L.~Peng, Y.~Cao, Y.~Sun, and Y.~Wang, ``Lightweight adaptive feature de-drifting for compressed image classification,'' \emph{IEEE Trans. Multimedia}, vol.~26, pp. 6424--6436, 2024.

\bibitem{peng2025boosting}
L.~Peng, Y.~Wang, X.~Di, X.~Fu, Y.~Cao, Z.-J. Zha \emph{et~al.}, ``Boosting image de-raining via central-surrounding synergistic convolution,'' in \emph{AAAI}, vol.~39, no.~6, 2025, pp. 6470--6478.

\bibitem{li2025fouriersr}
W.~Li, H.~Guo, Y.~Hou, and Z.~Ma, ``Fouriersr: A fourier token-based plugin for efficient image super-resolution,'' \emph{arXiv preprint arXiv:2503.10043}, 2025.

\bibitem{zamir2022restormer}
S.~W. Zamir, A.~Arora, S.~Khan, M.~Hayat, F.~S. Khan, and M.-H. Yang, ``Restormer: Efficient transformer for high-resolution image restoration,'' in \emph{CVPR}, 2022, pp. 5728--5739.

\bibitem{martin2001database}
D.~Martin, C.~Fowlkes, D.~Tal, and J.~Malik, ``A database of human segmented natural images and its application to evaluating segmentation algorithms and measuring ecological statistics,'' in \emph{ICCV}, vol.~2.\hskip 1em plus 0.5em minus 0.4em\relax IEEE, 2001, pp. 416--423.

\bibitem{huang2015single}
J.-B. Huang, A.~Singh, and N.~Ahuja, ``Single image super-resolution from transformed self-exemplars,'' in \emph{CVPR}, 2015, pp. 5197--5206.

\bibitem{ahn2018fast}
N.~Ahn, B.~Kang, and K.-A. Sohn, ``Fast, accurate, and lightweight super-resolution with cascading residual network,'' in \emph{ECCV}, 2018, pp. 252--268.

\bibitem{timofte2017ntire}
R.~Timofte, E.~Agustsson, L.~Van~Gool, M.-H. Yang, and L.~Zhang, ``Ntire 2017 challenge on single image super-resolution: Methods and results,'' in \emph{CVPRW}, 2017, pp. 114--125.

\bibitem{cai2019toward}
J.~Cai, H.~Zeng, H.~Yong, Z.~Cao, and L.~Zhang, ``Toward real-world single image super-resolution: A new benchmark and a new model,'' in \emph{ICCV}, 2019, pp. 3086--3095.

\bibitem{wang2002universal}
Z.~Wang and A.~C. Bovik, ``A universal image quality index,'' \emph{IEEE Signal Process. Lett.}, vol.~9, no.~3, pp. 81--84, 2002.

\bibitem{zhang2018unreasonable}
R.~Zhang, P.~Isola, A.~A. Efros, E.~Shechtman, and O.~Wang, ``The unreasonable effectiveness of deep features as a perceptual metric,'' in \emph{CVPR}, 2018, pp. 586--595.

\bibitem{chu2022improving}
X.~Chu, L.~Chen, C.~Chen, and X.~Lu, ``Improving image restoration by revisiting global information aggregation,'' in \emph{ECCV}.\hskip 1em plus 0.5em minus 0.4em\relax Springer, 2022, pp. 53--71.

\bibitem{gao2022lightweight}
G.~Gao, Z.~Wang, J.~Li, W.~Li, Y.~Yu, and T.~Zeng, ``Lightweight bimodal network for single-image super-resolution via symmetric cnn and recursive transformer,'' in \emph{IJCAI}, 2022, pp. 913--919.

\bibitem{yao2022wave}
T.~Yao, Y.~Pan, Y.~Li, C.-W. Ngo, and T.~Mei, ``Wave-vit: Unifying wavelet and transformers for visual representation learning,'' in \emph{ECCV}.\hskip 1em plus 0.5em minus 0.4em\relax Springer, 2022, pp. 328--345.

\bibitem{peng2024towards}
L.~Peng, W.~Li, R.~Pei, J.~Ren, J.~Xu, Y.~Wang, Y.~Cao, and Z.-J. Zha, ``Towards realistic data generation for real-world super-resolution,'' in \emph{ICLR}, 2025.

\bibitem{peng2025pixel}
L.~Peng, A.~Wu, W.~Li, P.~Xia, X.~Dai, X.~Zhang, X.~Di, H.~Sun, R.~Pei, Y.~Wang \emph{et~al.}, ``Pixel to gaussian: Ultra-fast continuous super-resolution with 2d gaussian modeling,'' \emph{arXiv preprint arXiv:2503.06617}, 2025.

\bibitem{jeevan2024wavemixsr}
P.~Jeevan, A.~Srinidhi, P.~Prathiba, and A.~Sethi, ``Wavemixsr: Resource-efficient neural network for image super-resolution,'' in \emph{WACV}, 2024, pp. 5884--5892.

\end{thebibliography}

\end{document}